\documentclass[	paper=a4, 
                fontsize=11pt, 
                DIV=11
                ]{scrartcl}
\usepackage[]{amsmath}
\usepackage{pdfsync}
\usepackage{microtype}

\makeatletter
\@ifundefined{XeTeXversion}{
\pdfoutput=1
\usepackage[pdftex]{graphicx}
\usepackage[T1]{fontenc}
\usepackage{paratype}
\usepackage[charter,cal=cmcal]{mathdesign}

}{%
\usepackage[xetex]{graphicx}
\usepackage[no-math]{fontspec}%
\setsansfont[Mapping=tex-text]{Calibri}
\setromanfont[Mapping=tex-text]{Cambria}
\usepackage{xunicode}%
\usepackage{xltxtra}%
\usepackage[xetex,
            colorlinks=true,
            linkcolor=spot,
            allcolors=spot,
            pdfauthor={Robin Ince},
            pdftitle={ }]{hyperref}
}
\makeatother
\graphicspath{{figs/}}
\usepackage[utf8]{inputenc}

\usepackage{setspace}
\recalctypearea

\usepackage[hyperref=true,
            backend=biber,
            bibencoding=utf8,
            style=authoryear,
            citestyle=authoryear,
            url=false,
            maxcitenames=2,
            maxbibnames=99,
            uniquename=false,
            isbn=false]{biblatex}

\usepackage[pdftex,
            colorlinks=true,
            linkcolor=spot,
            allcolors=spot,
            pdfauthor={Robin Ince},
            pdftitle={ }]{hyperref}

\DeclareFieldFormat{citehyperref}{%
  \DeclareFieldAlias{bibhyperref}{noformat}
  \bibhyperref{#1}}

\DeclareFieldFormat{textcitehyperref}{%
  \DeclareFieldAlias{bibhyperref}{noformat}
  \bibhyperref{%
    #1%
    \ifbool{cbx:parens}
      {\bibcloseparen\global\boolfalse{cbx:parens}}
      {}}}

\savebibmacro{cite}
\savebibmacro{textcite}

\renewbibmacro*{cite}{%
  \printtext[citehyperref]{%
    \restorebibmacro{cite}%
    \usebibmacro{cite}}}

\renewbibmacro*{textcite}{%
  \ifboolexpr{
    ( not test {\iffieldundef{prenote}} and
      test {\ifnumequal{\value{citecount}}{1}} )
    or
    ( not test {\iffieldundef{postnote}} and
      test {\ifnumequal{\value{citecount}}{\value{citetotal}}} )
  }
    {\DeclareFieldAlias{textcitehyperref}{noformat}}
    {}%
  \printtext[textcitehyperref]{%
    \restorebibmacro{textcite}%
    \usebibmacro{textcite}}}

\bibliography{fromzotero.bib}

\usepackage{chngpage}
\usepackage{url}
\usepackage[squaren]{SIunits}
\usepackage{color}
\definecolor{spot}{rgb}{0,0.2,0.6} 
\usepackage{marginnote}
 
\providecommand{\abs}[1]{\lvert#1\rvert} 

\usepackage{bm}

\usepackage{nicefrac}
\usepackage{makecell}
\usepackage{cellspace}
\usepackage{array}

\usepackage[format=plain,indention=.5cm,small,bf]{caption}

\begin{document}

\author{Robin A. A. Ince\\
\textit{\small Institute of Neuroscience and Psychology}\\[-1.5mm]
\textit{\small University of Glasgow, UK}\\
{\normalsize\href{mailto:robin.ince@glasgow.ac.uk}{\texttt{robin.ince@glasgow.ac.uk}}}
}

\date{}

\title{The Partial Entropy Decomposition: Decomposing multivariate entropy and mutual information via pointwise common surprisal}

\begin{singlespace}
    \maketitle
\end{singlespace}

\begin{abstract}
    Obtaining meaningful quantitative descriptions of the statistical dependence within multivariate systems is a difficult open problem.
    Recently, the Partial Information Decomposition (PID) was proposed to decompose mutual information (MI) about a target variable into components which are redundant, unique and synergistic within different subsets of predictor variables.
    Here, we propose to apply the elegant formalism of the PID to multivariate entropy, resulting in a Partial Entropy Decomposition (PED).
    We implement the PED with an entropy redundancy measure based on pointwise common surprisal; a natural definition which is closely related to the definition of MI.
    We show how this approach can reveal the dyadic vs triadic generative structure of multivariate systems that are indistinguishable with classical Shannon measures.
    The entropy perspective also shows that misinformation is synergistic entropy and hence that MI itself includes both redundant and synergistic effects.
    We show the relationships between the PED and MI in two predictors, and derive two alternative information decompositions which we illustrate on several example systems.
    This reveals that in entropy terms, univariate predictor MI is not a proper subset of the joint MI, and we suggest this previously unrecognised fact explains in part why obtaining a consistent PID has proven difficult. 
    The PED also allows separate quantification of mechanistic redundancy (related to the function of the system) versus source redundancy (arising from dependencies between inputs); an important distinction which no existing methods can address.
    The new perspective provided by the PED helps to clarify some of the difficulties encountered with the PID approach and the resulting decompositions provide useful tools for practical data analysis across a wide range of application areas. 

\end{abstract}



\section{Introduction}

Information theory was originally developed as a tool to study man-made communication systems \parencite{shannon_mathematical_1948,cover_elements_1991}. 
However, it also provides a general, unifying framework for many types of statistical analysis.
It has recently been suggested that most information theoretic measures are not capable of fully describing the dependency structure within multivariate systems in a meaningful way \parencite{james_multivariate_2016}. 
``Therefore, such measures are inadequate for discovering intrinsic causal relations'' \parencite{james_multivariate_2016}, an important problem in the study of complex systems as well as in many areas of experimental sciences (neuroscience, systems biology etc.).

\textcite{james_multivariate_2016} introduce two simple example systems which have equal values for all classical information theoretic measures --- i.e. equal \emph{I-diagrams} \parencite{yeung_new_1991} --- but completely different generative structure.
They show very few existing measures can distinguish between these two systems, and even those that can do not clearly reflect the different generative structures.
The challenge they pose is to allocate the $3$ bits of entropy in these two systems in a way that meaningfully reflects the different structures of dependency.
They suggest this requires ``a fundamentally new measure''.

Here, we propose a new measure which we believe answers the challenge of \textcite{james_multivariate_2016}.
\textcite{williams_nonnegative_2010} developed an elegant mathematical framework called the Partial Information Decomposition(PID).
The PID decomposes mutual information between a target variable and a multivariate set of predictor variables into a set of terms quantifying information which arises uniquely, redundantly or synergistically within or between different subsets of predictors \parencite{timme_synergy_2013}.
We propose applying the framework of the PID directly to multivariate entropy rather than to multivariate mutual information about a chosen target variable.
The concepts of redundancy (shared information) and synergy (additional information arising when variables are considered together) can be applied directly to entropy.
Redundant entropy is uncertainty that is common to a set of variables; synergistic entropy is additional uncertainty that arises when variables are combined.

Here, we develop the consequences of this Partial Entropy Decomposition (PED), implementing it with an entropy redundancy measure based on pointwise common surprisal \parencite{ince_measuring_2016}.
We show this perspective provides new insights into negative local information values, or misinformation \parencite{wibral_bits_2014} --- these terms quantify synergistic entropy. 
It also sheds light on some of the difficulties around obtaining a consistent PID; even the axioms required for a meaningful information redundancy measure have been difficult to pin down and are currently debated \parencite{harder_bivariate_2013,bertschinger_quantifying_2014,griffith_quantifying_2014,griffith_intersection_2014,rauh_reconsidering_2014-1,ince_measuring_2016,bertschinger_shared_2013,olbrich_information_2015,griffith_quantifying_2014,rauh_extractable_2017,banerjee_new_2014}.  

We first briefly review the Partial Information Decomposition, then introduce the Partial Entropy Decomposition and our proposed entropy redundancy measure.
We explore the relationships between the PED and mutual information, and use these to derive two different decompositions of multivariate mutual information.
We show how the PED can meaningfully quantify the example systems of \textcite{james_multivariate_2016} and compare the new information decompositions to existing PID approaches on a range of examples.

\section{Background}

\subsection{Partial Information Decomposition}

\textcite{williams_nonnegative_2010} propose a non-negative decomposition of mutual information conveyed by a set of predictor variables $\mathcal{X}=\left\{ X_1, X_2, \dots, X_n \right\}$, about a target variable $S$. 
They reduce the total multivariate mutual information, $I(\mathcal{X}; S)$, into a number of atoms representing the unique, redundant and synergistic information between all subsets of $\mathcal{X}$.
To do this they consider all subsets of $\mathcal{X}$, denoted $\mathbf{A_i}$, and termed \emph{sources}.
They show that the redundancy structure of the multi-variate information is determined by the ``collection of all sets of sources such that no source is a superset of any other'' --- formally the set of antichains on the lattice formed from the power set of $\mathcal{X}$ under set inclusion, denoted $\mathcal{A}(\mathcal{X})$.
Together with a natural ordering, this defines an information redundancy lattice \parencite{crampton_completion_2001}.
Each node of the lattice represents a partial information atom, the value of which is given by a partial information (PI) function.

\begin{figure}[htbp]
    \centering
    \includegraphics[width=0.8\textwidth]{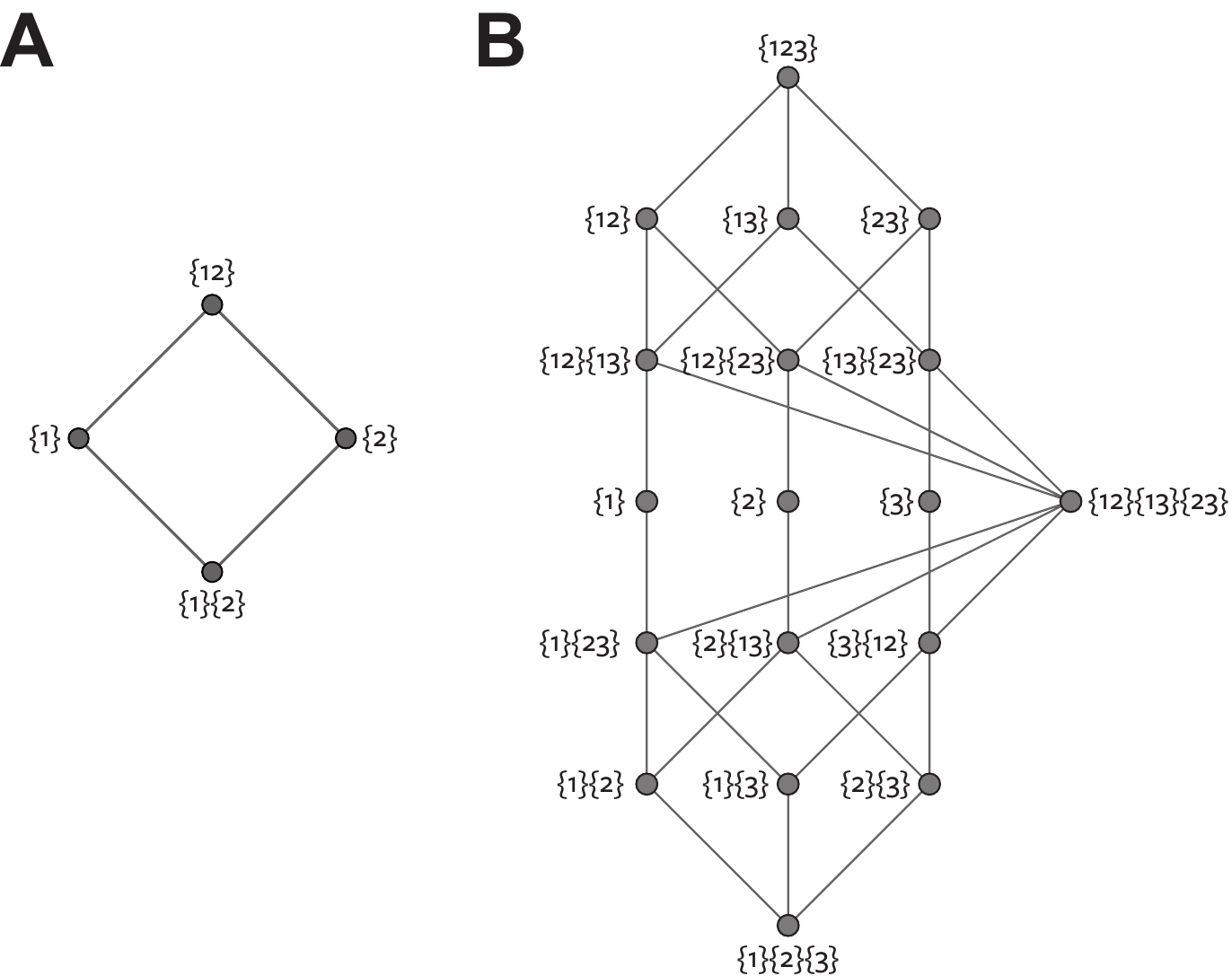}
    \caption{\emph{Redundancy lattice} for 
    \textbf{A.} two variables, 
    \textbf{B.} three variables. 
    Modified from \textcite{williams_nonnegative_2010}.}
    \label{fig:lattice}
\end{figure}

Figure \ref{fig:lattice} shows the structure of this lattice for $n=2,3$.
Each node is defined as a set of sources.
Sources can be multivariate and are indicated with braces.
We drop the explicit $X$ notation so $\{1\}$ corresponds to a source containing the variable $X_1$.
When a node contains multiple sources, it indicates the redundancy or common information between those sources, so $\{1\}\{2\}$ represents the information common to or shared between $X_1$ and $X_2$.
The information redundancy value of each node of the lattice, $I_\cap$, measures the total information provided by that node; the partial information function, $I_\partial$, measures the unique information contributed by only that node (redundant, synergistic or unique information within or between subsets of variables).
So the information redundancy of node $\{12\}$ is the joint mutual information $I(S;X_1,X_2)$, and the partial information value of that node represents the synergy, that is information about $S$ that arises only when $X_1$ and $X_2$ are considered together.
The PI value for each node can be determined via a recursive relationship (M\"obius inverse) over the information redundancy values of the lattice:
\begin{align}
    I_\partial(S; \alpha) = I_\cap (S; \alpha) - \sum_{\beta \preceq \alpha} I_\cap (S; \beta)
    \label{eq:mobius}
\end{align}
where $\alpha \in \mathcal{A}(\mathcal{X})$ is a set of sources (each a set of input variables $X_i$) defining the node in question.

Note that in the two variable PID, the bivariate mutual information values are related to the partial information terms as follows:
\begin{align}
    I(S; X_1,X_2) = I_\partial(S;\{1\}\{2\}) + I_\partial(S;\{1\}) + I_\partial(S;\{2\})+ I_\partial(S;\{1 2\}) 
    \label{eq:mi12pid}
\end{align}
Thus, the bivariate mutual information is split into four terms representing a contribution shared between both variables, unique to each variable and synergistic between the two.
Similarly the individual mutual information values are decomposed as the sum of the shared and the unique contributions.
\begin{align}
\begin{split}
    I(S; X_1) &= I_\partial(S;\{1\}\{2\}) + I_\partial(S;\{1\}) \\
    I(S; X_2) &= I_\partial(S;\{1\}\{2\}) + I_\partial(S;\{2\}) 
    \label{eq:mi1pid}
\end{split}
\end{align}

\section{The Partial Entropy Decomposition}

Here we apply the mathematical framework of the PID \parencite{williams_nonnegative_2010} directly to multivariate entropy, to obtain a Partial Entropy Decomposition (PED). 
Consider the lattices shown in Figure 1.
The terms are the same as the PID, but now there is no target variable and they represent the \emph{entropy} that is shared (redundant), unique or synergistic between the variables.

We denote the overall entropy of each node of the lattice as $H_\cap$; this measures the total entropy provided by that node. We then define a partial entropy function, $H_\partial$, defined analogously to $I_\partial$ via M\"obius inversion (Eq.~\ref{eq:mobius}), which measures the unique entropy contributed by only that node (redundant, synergistic or unique entropy within subsets of variables).

For the two variable case, if the redundancy function used for a set of sources is denoted $H_\cap\left(\mathbf{A_1},\dots,\mathbf{A_k}\right)$ and following the notation in \textcite{williams_nonnegative_2010,ince_measuring_2016}, the nodes of the lattice, their entropy redundancy and their partial entropy values are given in Table \ref{tab:2varterms}.

\begin{table}[htbp]
\centering\setcellgapes{4pt}\makegapedcells \renewcommand\theadfont{\normalsize\bfseries}
    \begin{tabular}{| c | p{3.7cm} | p{2.7cm} | p{4.3cm} |}
        \hline
        \thead{Node label} & \thead{Redundancy \\ function} & \thead{Partial entropy \\ $H_\partial$} & \thead{Represented atom} \\
        \hline \hline
        \{12\} & $H_\cap(\{ 1 2 \} ) = H(X_1,X_2)$ & $H_\cap(\{1 2\})$ \newline -$H_\cap(\{1\} )$ \newline - $H_\cap(\{ 2\})$ \newline +$H_\cap(\{ 1 \} \{ 2 \} )$   & entropy available only from \newline $X_1$ and $X_2$ together \newline (synergy) \\ \hline
        \{1\} & $H_\cap(\{1\})=H(X_1)$ & $H_\cap(\{ 1 \} )$ \newline - $H_\cap(\{ 1 \} \{ 2 \} )$ &  unique entropy in $X_1$ only \\ \hline
        \{2\} & $H_\cap(\{2\})=H(X_2)$ & $H_\cap(\{ 2 \} )$ \newline - $H_\cap(\{ 1 \} \{ 2 \} )$ & unique entropy in $X_2$ only \\ \hline
        \{1\}\{2\} & $H_\cap(\{1\}\{2\})$ & $H_\cap(\{1\} \{2\})$ & entropy shared between \newline $X_1$ and $X_2$ \\ \hline
  \end{tabular}
  \caption{\emph{Full PED in the two-variable case.}}
  \label{tab:2varterms}
\end{table}

Again, in direct analogy with the PID we have:

\begin{align}
    \begin{split}
    H(X_1) &= H_\partial (\{1\}\{2\}) + H_\partial (\{1\}) \\
    H(X_2) &= H_\partial (\{1\}\{2\}) + H_\partial (\{2\}) \\
    H(X_1,X_2) &= H_\partial (\{1\}\{2\}) + H_\partial (\{1\}) + H_\partial (\{2\}) + H_\partial (\{12\})
\end{split}
\end{align}

Inserting these partial entropy values into the definition of mutual information we see that

\begin{align}
    I(X_1;X_2) &= H_\partial(\{1\}\{2\}) - H_\partial(\{12\}) 
    \label{eq:mi_ped}
\end{align}

That is, mutual information is redundant or shared entropy minus synergistic entropy.
Therefore, in the same way that interaction information conflates redundant and synergistic information, mutual information itself combines redundant and synergistic entropy.
We suggest that this is an important shift of perspective that helps clarify many of the difficulties observed in defining multivariate redundancy measures for mutual information and obtaining consistent PIDs.

In fact Eq. \ref{eq:mi_ped} suggests a direct interpretation of negative local or pointwise information terms, which have been called \emph{misinformation} \parencite{wibral_bits_2014}.
While the partial entropy terms depend on the particular entropy redundancy measure used, it suggests that positive local terms correspond to redundant entropy while negative local misinformation terms correspond to synergistic entropy (this is explicitly the case for the common surprisal entropy redundancy measure we propose below).
Therefore, \emph{misinformation is synergistic entropy}.
This makes intuitive sense, misinformation means an observer is less likely to see those two particular values together than would be expected if they were independent.
Alternatively, there is an increase in local surprisal when observing both variables together, versus observing each independently.
Therefore, this local term contributes to an increase in entropy when both variables are observed together --- which is by definition synergistic entropy.

Even before defining an entropy redundancy function, this perspective can give insights into problems faced while defining axioms that an information redundancy measure should satisfy.
For example, the identity property has been proposed as one such axiom \textcite{harder_bivariate_2013}:
\begin{equation}
    I_\cap\left(\{\mathbf{A_1},\mathbf{A_2}\}; \mathbf{A_1},\mathbf{A_2}\right) = I(\mathbf{A_1}; \mathbf{A_2})
    \label{eq:idaxiom}
\end{equation}

By considering the PED (Eq. \ref{eq:mi_ped}) we can see that if there is synergistic entropy between $A_1$ and $A_2$ then this will feature in the right hand side mutual information term.
However, this should not appear in the $I_\cap$ term on the left hand side since the redundancy between $A_1$ and $A_2$ should not include synergistic entropy which occurs only when $A_1$ and $A_2$ are observed together. 
This suggests that the identity axiom should instead be defined as:
\begin{equation}
    I_\cap\left(\{\mathbf{A_1},\mathbf{A_2}\}; \mathbf{A_1},\mathbf{A_2}\right) = H_\cap(\{\mathbf{A_1}\}\{\mathbf{A_2}\})
    \label{eq:modidaxiom}
\end{equation}

\subsection{Measuring shared entropy as common surprisal}

The following four axioms have been suggested for an information redundancy measure  \parencite{williams_nonnegative_2010,harder_bivariate_2013,ince_measuring_2016}, and we propose they should be considered also for an entropy redundancy measure:

\vspace{0.3cm}
\textbf{Symmetry:} 
\begin{equation}
    H_\cap\left(\mathbf{A_1},\dots,\mathbf{A_k} \right) \text{is symmetric with respect to the } \mathbf{A_i}\text{'s.}
\end{equation}

\textbf{Self Redundancy:} 
\begin{equation}
    H_\cap\left(\mathbf{A} \right) = H(\mathbf{A})
\end{equation}

\textbf{Subset Equality:} 
\begin{equation}
    H_\cap\left(\mathbf{A_1},\dots,\mathbf{A_{k-1}},\mathbf{A_k} \right) = H_\cap\left(\mathbf{A_1},\dots,\mathbf{A_{k-1}} \right)
    \mbox{ if } \mathbf{A_{k-1}} \subseteq \mathbf{A_k}
\end{equation}

\textbf{Monotonicity:} 
\begin{equation}
    H_\cap\left(\mathbf{A_1},\dots,\mathbf{A_{k-1}},\mathbf{A_k} \right) \leq H_\cap\left(\mathbf{A_1},\dots,\mathbf{A_{k-1}} \right)
\end{equation}

Note that we have separated subset equality from the monotonicity axiom \parencite{ince_measuring_2016}, since subset equality is the condition required for the redundancy lattice to be a complete description of the dependency structure of the system, independent of whether the measure is monotonic on the lattice.

Many information theoretic quantities can be formulated as the expectation over a multivariate distribution of a particular functional. 
The terms over which the expectation is taken, the value of the functional for specific values of the considered variable, are called local or pointwise terms \parencite{wibral_local_2014,lizier_local_2008,wibral_bits_2014,van_de_cruys_two_2011,church_word_1990}. 
For example, entropy is defined as the expectation of surprisal:
\begin{equation}
    H(X) = \sum_{x \in X} p(x) h(x)
\end{equation}
Here the surprisal, $h(x) = -\log p(x)$, is the local or pointwise functional corresponding to the entropy measure. 

In \textcite{ince_measuring_2016} we present an information redundancy measure based on pointwise common change in surprisal (since local information is a change in surprisal).
Here, we apply the same principles to define a measure of entropy redundancy based on pointwise common surprisal.
We define the common surprisal, $H_\text{cs}$, as the sum of positive pointwise co-information values \parencite{bell_co-information_2003,matsuda_physical_2000,jakulin_quantifying_2003,timme_synergy_2013}. 
While \textcite{ince_measuring_2016} motivated $I_\text{ccs}$ in terms of interaction information \parencite{mcgill_multivariate_1954}, we use co-information here.
Co-information and interaction information are equivalent in magnitude, but have opposite signs over odd numbers of variables. 
In the context of the well known set-theoretical interpretation of information theoretic quantities as measures which quantify the area of sets and which can be visualised with Venn diagrams \parencite{reza_introduction_1961}, co-information can be derived as the set intersection of multiple entropy values. 
This set theoretic interpretation can also be applied at the local level, so local co-information measures the set-theoretic overlap of multiple local entropy values; hence common or shared surprisal.
Local co-information can be negative, but since the local informations over which the intersection is calculated are positive in this case there is no overlap. 
Note that for the information measure $I_\text{ccs}$, care had to be taken regarding the different signs of the change in surprisals (local information values) that the overlap was being calculated over, but here each input surprisal value is always positive. 
For a redundancy function we want to measure only positive overlap so we simply ignore the negative values.
\begin{align}
    H_\text{cs} (\mathbf{A_1}, \dots, \mathbf{A_n}) &= \sum_{a_1,\dots,a_n} p(a_1,\dots, a_n) h_\text{cs}(a_1,\dots,a_n) \label{eq:hcsdefexp}\\
    h_\text{cs}(a_1,\dots,a_n) &= \max\left[ c(a_1,\dots,a_n), 0 \right] 
\end{align}
where $c(a_1,\dots,a_n)$ is the local co-information, defined as \parencite{matsuda_physical_2000}: 
\begin{align}
    c(a_1,\dots,a_n) = \sum_{k=1}^n (-1)^{k+1} \sum_{i_1 < \dots < i_k} h\left(a_{i_1},\dots,a_{i_k}\right)
    \label{eq:coinfo}
\end{align}
and $h(\mathbf{a}) = - \log_2 p(\mathbf{a})$ is the surprisal (local entropy).

For two variables, we can see that as suggested by Eq.~\ref{eq:mi_ped}, this definition provides that redundant entropy is exactly the sum of positive local mutual information terms, while synergistic entropy is the sum of negative local mutual information, or misinformation terms. 

For three variable redundancy, the definition of co-information requires a joint distribution over the three sources.
Analogous to the arguments for why an information redundancy measure should depend only on the pairwise target-source marginals \parencite{ince_measuring_2016,bertschinger_quantifying_2014}, we suggest that three way shared entropy should depend only on pairwise marginals and hence use the maximum entropy distribution that preserves two-way marginals.

The symmetry and self-redundancy axioms are satisfied from the properties of co-information \parencite{matsuda_physical_2000}.
Subset-equality is also satisfied.
If $\mathbf{A_{l-1}} \subseteq \mathbf{A_l}$ then we consider values
$a_{l-1} \in \mathbf{A_{l-1}}$, $a_{l} \in \mathbf{A_{l}}$ with $a_l = (a_l^{l-1},a_l^+)$ and $a_l^{l-1} \in \mathbf{A_{l-1}} \cap \mathbf{A_l} = \mathbf{A_{l-1}}$, $a_l^+ \in \mathbf{A_l} \setminus \mathbf{A_{l-1}}$. 
Then
\begin{align}
\begin{split}
    p(a_{i_1},\dots,a_{i_j},a_{l-1},a_l^{l-1},a_l^+) = 
    \begin{cases}
        0 &\text{if } a_{l-1} \neq a_l^{l-1} \\
        p(a_{i_1},\dots,a_{i_j},a_l) &\text{otherwise}   
\end{cases}
\end{split}
\end{align}
for any $i_1<\dots<i_j \in \{1,\dots,l-2\}$.
So for non-zero terms in Eq.~\ref{eq:hcsdefexp}:
\begin{equation}
h(a_{i_1},\dots,a_{i_j},a_{l-1},a_l) = h(a_{i_1},\dots,a_{i_j},a_l) 
\end{equation}
 
Therefore all terms for $k \geq 2$ in Eq.~\ref{eq:coinfo} which include $a_{l-1},a_l$ cancel with a corresponding $k-1$ order term including $a_l$, so
\begin{equation}
    c(a_1,\dots,a_{l-1},a_l) = c(a_1,\dots,a_{l-1})
\end{equation}
and subset equality holds.

However, as we will see in the examples, monotonicity is not satisfied.
$H_\text{cs}$ is continuous in $p$ by the continuity of surprisal.
It is a linear combination of surprisal values thresholded with 0, therefore it is continuous but not differentiable.

\subsection{Relating a 3 variable PED to 2 variable PEDs}

\begin{figure}[htbp]
    \centering
    \includegraphics[width=0.8\textwidth]{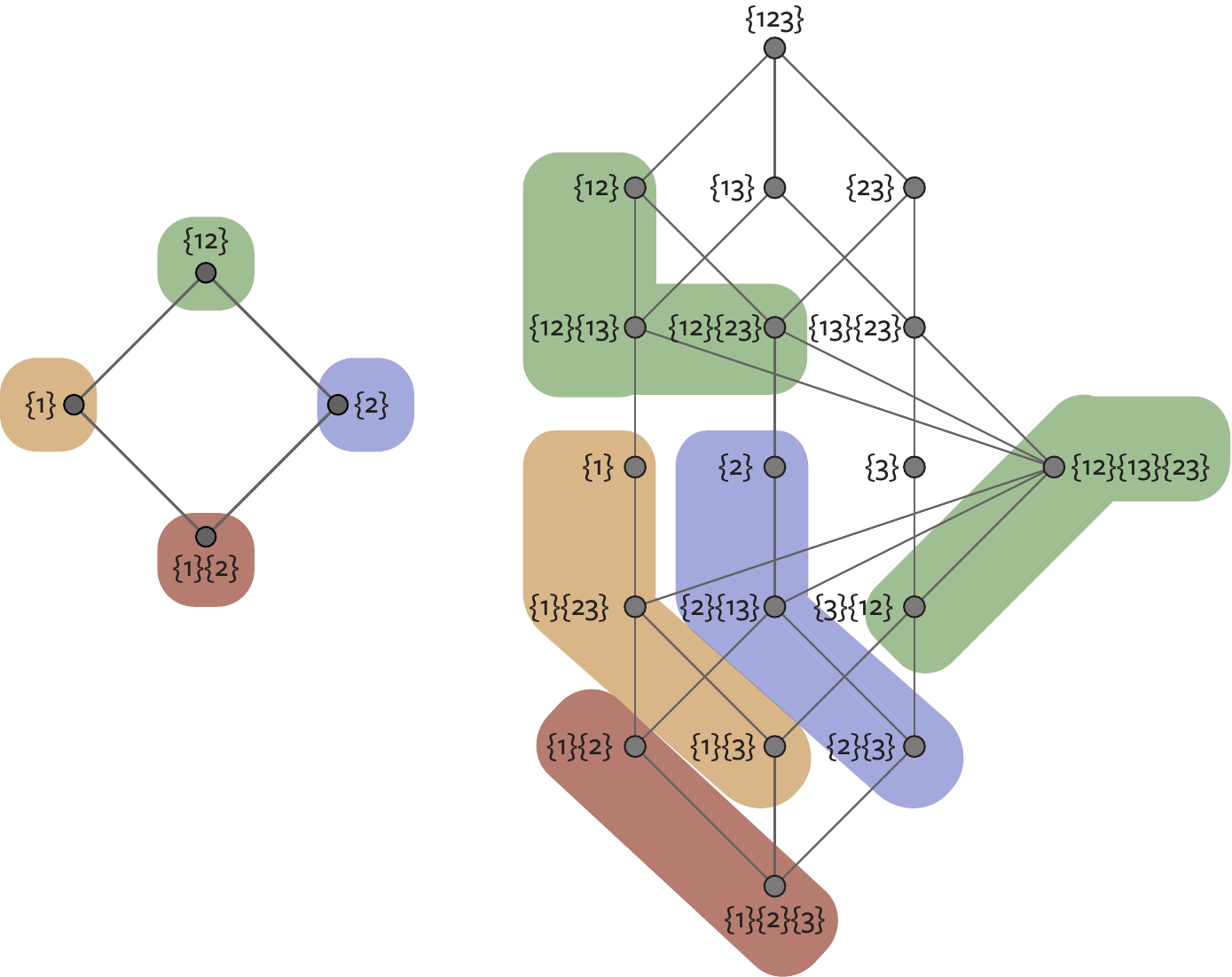}
    \caption{\emph{Relations between 3 variables and marginalised 2 variable entropy redundancy lattices} }
    \label{fig:ped2v3}
\end{figure}

We can consider the effect of marginalising out a variable from a three variable system to obtain a two variable system.
When marginalising away a variable, the two variable PED terms should be the sum of all three variable terms including the same sources.
These equality relationships detailed in Table~\ref{tab:ped2v3} and illustrated by the colored terms in Figure \ref{fig:ped2v3}.
The same relationships are noted for information lattices in \parencite{chicharro_synergy_2017}.
To distinguish between different partial entropy values we denote the space over which the PED is considered with a superscript.
In the marginalised PED it does not matter that the partial entropy $H^{123}_\partial(\{1\}\{23\})$ is commonly available from $\{1\}$ and synergistically from $\{23\}$. 
When $X_3$ is removed it is unique entropy in $H^{12}_\partial(\{1\})$.
These relations come from the structure of the lattice and so should be met by any consistent entropy redundancy function. 

\begin{table}[htbp]
\centering\setcellgapes{4pt}\makegapedcells \renewcommand\theadfont{\normalsize\bfseries}%
 \begin{tabular}{|l| l|}
 \hline
 \thead{Marginalised\\ term} & \thead{3 variable terms} \\
 \hline \hline
 $H_\partial^{12}(\{12\})$ &
 $\begin{aligned}
 &H_\partial^{123}(\{12\}) + H_\partial^{123}(\{3\}\{12\}) \\
 &+ H_\partial^{123}(\{12\}\{13\}) + H_\partial^{123}(\{12\}\{23\}) \\
 &+ H_\partial^{123}(\{12\}\{13\}\{23\})
 \end{aligned}$ \\
 \hline
 $H_\partial^{12}(\{1\})$ &
 $H_\partial^{123}(\{1\})+H_\partial^{123}(\{1\}\{23\})+H_\partial^{123}(\{1\}\{3\})$ \\
 \hline
 $H_\partial^{12}(\{2\})$ &
 $H_\partial^{123}(\{2\})+H_\partial^{123}(\{2\}\{13\})+H_\partial^{123}(\{2\}\{3\})$ \\
 \hline
 $H_\partial^{12}(\{1\}\{2\})$ &
 $H_\partial^{123}(\{1\}\{2\})+H_\partial^{123}(\{1\}\{2\}\{3\})$ \\
 \hline
 \end{tabular}
   \caption{\emph{Marginalising over variable $X_3$.}}
  \label{tab:ped2v3}
\end{table}

\begin{figure}[htbp]
    \centering
    \includegraphics[width=0.8\textwidth]{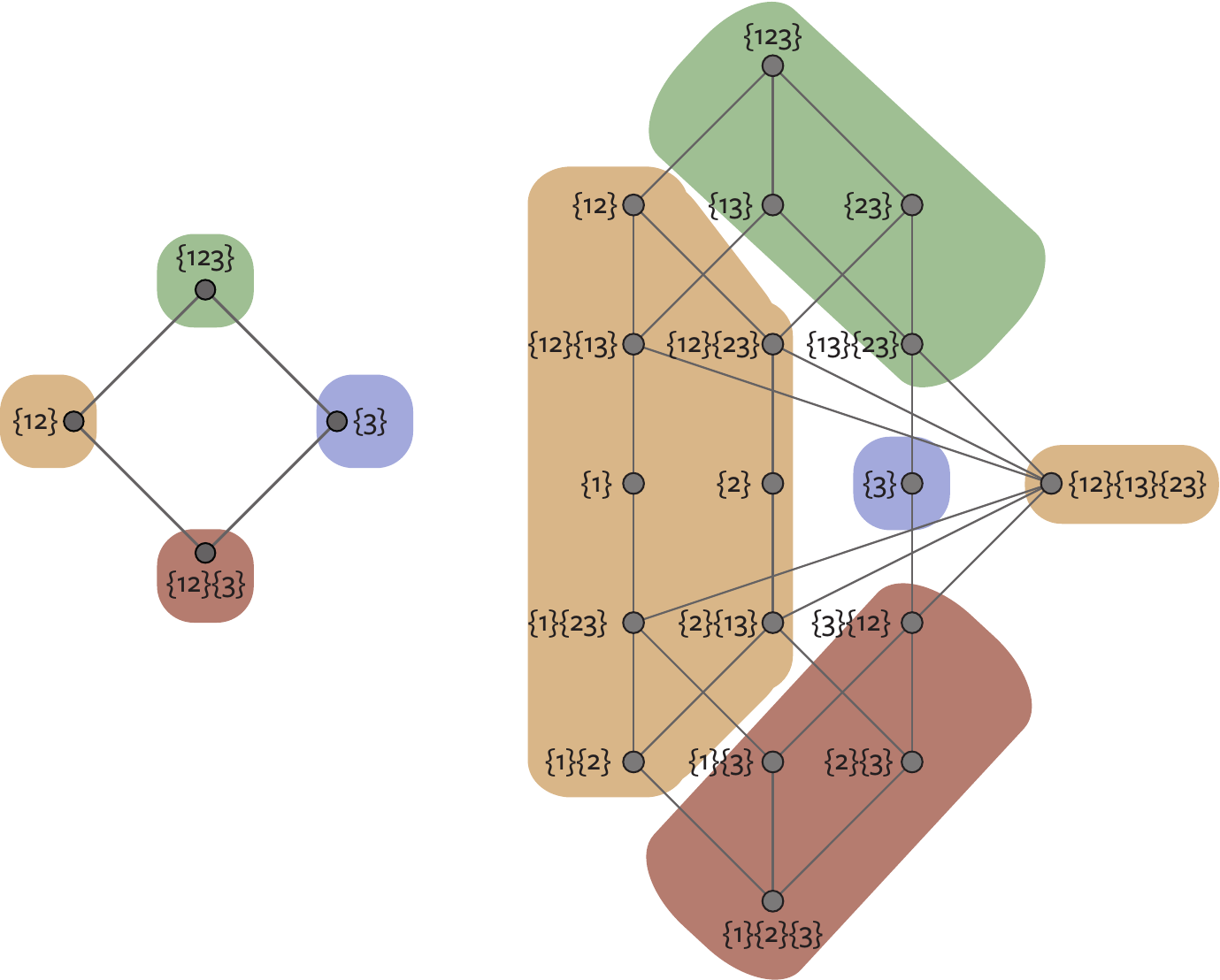}
    \caption{\emph{Relations between entropy redundancy lattice for three variables considered separately, and the bivariate lattice when $X_1$, $X_2$ are considered as a single combined variable.} }
    \label{fig:ped12v3}
\end{figure}

We can also consider the two variable PID obtained from combining $X_1$ and $X_2$ into a single variable.
These equality relationships detailed in Table~\ref{tab:ped12v3} and illustrated by the colored terms in Figure \ref{fig:ped12v3}.
As an example although the variable $X_3$ appears in the term $H^{123}_\partial(\{1\}\{23\})$ it is in a synergistic combination.
$H^{123}_\partial(\{1\}\{23\})$ quantifies entropy that is available synergistically between $\{23\}$, but also in $\{1\}$.
In the context of variables $X_1$ and $X_2$ being combined this means that partial entropy is not shared between $X_1$ and $X_2$ (since it arises synergistically when $X_2$ is combined with $X_3$). 
It is available when all three variables are considered together, but not uniquely, since variable $X_1$ alone already contains it. 
It therefore belongs in the $H^{(12)3}_\partial(\{12\})$ term.

\begin{table}[htbp]
\centering\setcellgapes{4pt}\makegapedcells \renewcommand\theadfont{\normalsize\bfseries}
    \begin{tabular}{| l | l |}
        \hline
        \thead{Marginalised term} & \thead{3 variable terms} \\
        \hline \hline 
        $H_\partial^{(12)3}(\{123\})$ &  
        $H_\partial^{123}(\{123\})+H_\partial^{123}(\{13\})+H_\partial^{123}(\{23\})+H_\partial^{123}(\{12\}\{23\})$ \\
        \hline
        $H_\partial^{(12)3}(\{12\})$ & 
        $\begin{aligned} 
            &H_\partial^{123}(\{12\}) + H_\partial^{123}(\{12\}\{13\}) + H_\partial^{123}(\{12\}\{23\}) \\
            &+ H_\partial^{123}(\{1\})+ H_\partial^{123}(\{2\})  + H_\partial^{123}(\{12\}\{13\}\{23\}) \\ 
            &+ H_\partial^{123}(\{1\}\{23\}) + H_\partial^{123}(\{2\}\{13\})+ H_\partial^{123}(\{1\}\{2\}) 
            \end{aligned}$ \\
        \hline
        $H_\partial^{(12)3}(\{3\})$ &  
        $H_\partial^{123}(\{3\}) $\\
        \hline
        $H_\partial^{(12)3}(\{12\}\{3\})$ &  
        $H_\partial^{123}(\{3\}\{12\}) + H_\partial^{123}(\{1\}\{3\})+ H_\partial^{123}(\{2\}\{3\}) + H_\partial^{123}(\{1\}\{2\}\{3\})$ \\
        \hline
  \end{tabular}
  \caption{\emph{Combining variables $X_1$ and $X_2$.}}
  \label{tab:ped12v3}
\end{table}

\subsection{Mutual information from PED terms}
\label{sec:mifromped}

The Partial Information Decomposition considers information about a target variable, between two or more predictor variables.
Starting from the simplest case with two predictor variables, this involves a three variable system.
From Eq.~\ref{eq:mi_ped} and Table~\ref{tab:ped2v3} we can see:
\begin{align}
\begin{split}
    I(X_1;X_3) &= H^{13}_\partial(\{1\}\{3\}) - H^{13}_\partial(\{13\}) \\
    &= H^{123}_\partial(\{1\}\{3\}) + H^{123}_\partial(\{1\}\{2\}\{3\}) \\
    &\phantom{{}={}} - H^{123}_\partial(\{13\}) - H^{123}_\partial(\{12\}\{13\}) 
    - H^{123}_\partial(\{13\}\{23\}) \\ 
    &\phantom{{}={}} - H^{123}_\partial(\{2\}\{13\}) - H^{123}_\partial(\{12\}\{13\}\{23\}) 
\end{split}
\label{eq:mi1}
\end{align}
and similarly (from here all $H_\partial$ terms are in the three variable space so we drop the superscript):
\begin{align}
\begin{split}
    I(X_2;X_3) &=  H_\partial(\{2\}\{3\}) + H_\partial(\{1\}\{2\}\{3\}) \\
    &\phantom{{}={}} - H_\partial(\{23\}) - H_\partial(\{12\}\{23\}) 
    - H_\partial(\{13\}\{23\}) \\ 
    &\phantom{{}={}} - H_\partial(\{1\}\{23\}) - H_\partial(\{12\}\{13\}\{23\}) 
\end{split}
\label{eq:mi2}
\end{align}

Similarly, from Eq.~\ref{eq:mi_ped} and Table~\ref{tab:ped12v3} we have:
\begin{align}
\begin{split}
    I(X_1,X_2;X_3) &=  H_\partial(\{3\}\{12\}) + H_\partial(\{1\}\{3\}) + H_\partial(\{2\}\{3\}) + H_\partial(\{1\}\{2\}\{3\})\\
    &\phantom{{}={}} - H_\partial(\{123\}) - H_\partial(\{13\})- H_\partial(\{23\})  
    - H_\partial(\{13\}\{23\})
\end{split}
\label{eq:mi12}
\end{align}

The conditional mutual information can also be written in terms of partial entropy values:
\begin{align}
\begin{split}
    I(X_1;X_3|X_2) &=  H_\partial(\{3\}\{12\}) + H_\partial(\{1\}\{3\}) - H_\partial(\{123\}) - H_\partial(\{13\})\\
    &\phantom{{}={}} + H_\partial(\{12\}\{23\}) + H_\partial(\{1\}\{23\}) + H_\partial(\{12\}\{13\}\{23\}) 
\end{split}
\label{eq:cmi}
\end{align}

\subsection{PID from PED}

In the previous section we obtained expressions for the bivariate mutual informations involved in the two variable partial information lattice in terms of three variable partial entropy terms.
With variable $X_3$ as the target, we can directly relate the two univariate mutual information partial entropy equations (Eqs.~\ref{eq:mi1},~\ref{eq:mi2}).
Separating the terms that are common to the two suggests the following partial information decomposition:
\begin{align}
\begin{split}
    I_\partial(3; \{1\}\{2\}) &= H_\partial(\{1\}\{2\}\{3\}) - H_\partial(\{13\}\{23\}) - H_\partial(\{12\}\{13\}\{23\}) \\ 
    I_\partial(3; \{1\}) &= H_\partial(\{1\}\{3\}) - H_\partial(\{13\}) - H_\partial(\{12\}\{13\}) - H_\partial(\{2\}\{13\}) \\ 
    I_\partial(3; \{2\}) &= H_\partial(\{2\}\{3\}) - H_\partial(\{23\}) - H_\partial(\{12\}\{23\}) - H_\partial(\{1\}\{23\}) \\ 
\end{split}
\label{eq:pedpid1}
\end{align}
Subtracting the sum of these terms from the the joint mutual information (Eq.~\ref{eq:mi12}) yields the following synergy term:
\begin{align}
\begin{split}
    I_\partial(3; \{12\}) &= H_\partial(\{3\}\{12\}) + H_\partial(\{12\}\{13\}) + H_\partial(\{12\}\{23\}) \\
     &\phantom{{}={}} + H_\partial(\{12\}\{13\}\{23\}) + H_\partial(\{1\}\{23\}) + H_\partial(\{2\}\{13\}) \\
     &\phantom{{}={}} - H_\partial(\{123\})
\end{split}
\label{eq:pedpid1syn}
\end{align}

This results in a valid PID that satisfies Eqs.~\ref{eq:mi12pid} and \ref{eq:mi1pid}.
However, it does not produce a non-negative decomposition (with the $H_\text{cs}$ measure proposed here).
As previously noted \parencite{ince_measuring_2016}, it is possible for a variable to convey unique misinformation (see example \textsc{ImperfectRdn}, Section \ref{sec:otherex}).
Here, partial entropy terms which are subtracted represent misinformation (since they include synergistic entropy between a predictor and the target variable).
While we derive this PID here from the relations between two and three variable entropy redundancy lattices, it is also possible to arrive at the same result simply by considering the interpretation of each partial entropy term directly. 
Using the insight that synergistic entropy is misinformation, we can see for example that $H_\partial(\{1\}\{3\})$ represents unique information about $X_3$ in $X_1$ while $H_\partial(\{13\})$ is unique misinformation about $X_3$ in $X_1$ and so should be subtracted from the unique partial information term. 
It is clear that $H_\partial(\{1\}\{2\}\{3\})$ represents redundant information (redundant entropy shared between all three variables) and $H_\partial(\{12\}\{3\})$ represents synergistic information (synergistic entropy between $X_1,X_2$ which is redundant with, therefore informative about, $X_3$).
$H_\partial(\{13\}\{23\})$ is misinformation (synergistic entropy) that is common to both $X_1$ and $X_2$, so is subtracted from the redundant partial information term. 

However, while these terms have clear interpretations, there are a number of partial entropy terms that are more ambiguous.
They appear twice in the PID with opposite signs, and indeed closer consideration shows they do indeed quantify entropy that is inherently ambiguous.
For example, $H_\partial(\{1\}\{23\})$ quantifies unique misinformation in $X_2$ in the context of $I(X_2;X_3)$ with $X_1$ marginalized out. 
But it also quantifies synergistic information in $X_1,X_2$ in the context of the PID of $I(X_1,X_2;X_3)$, since it quantifies redundant entropy between $X_1$ and $X_3$ (which appear in two redundant source terms), but which is only available when $X_2$ is also considered. 
When $X_1$ and $X_2$ are both available, the term quantifies entropy gained from the addition of $X_3$ ($\{23\}$), which overlaps with entropy in $X_1$. 
However, when $X_1$ and $X_2$ are considered as a combined variable as in Table~\ref{tab:ped2v3} the term is unique entropy in $\{12\}$ since it is fully available already from observing $(X_1,X_2)$.

Similarly $H_\partial(\{12\}\{13\})$ is both unique misinformation in the context of $I(X_1;X_3)$ as well as synergistic information in the context of a PID of $I(X_1,X_2;X_3)$.
$H_\partial(\{12\}\{13\}\{23\})$ is again synergistic information in the context of $I(X_1,X_2;X_3)$ but redundant misinformation in the context of $I(X_1;X_3)$ and $I(X_2;X_3)$ because it quantifies redundant synergistic entropy terms. 

\subsection{An alternative PID ignoring ambiguous partial entropy terms}

The presence of the repeated ambiguous entropy terms in the partial entropy derived PID described in the previous section, together with Eq.~\ref{eq:mi12}, suggests an alternative decomposition of the mutual information $I(X_1,X_2;X_3)$ as follows. 
Since each term in Eq.~\ref{eq:mi12} admits an unambiguous interpretation we can define:
\begin{align}
\begin{split}
    I_\partial(3; \{1\}\{2\}) &= H_\partial(\{1\}\{2\}\{3\}) - H_\partial(\{13\}\{23\})  \\ 
    I_\partial(3; \{1\}) &= H_\partial(\{1\}\{3\}) - H_\partial(\{13\})  \\ 
    I_\partial(3; \{2\}) &= H_\partial(\{2\}\{3\}) - H_\partial(\{23\})  \\ 
    I_\partial(3; \{12\}) &= H_\partial(\{3\}\{12\})  - H_\partial(\{123\})
\end{split}
\label{eq:pedpid2}
\end{align}
Since this approach is obtained by omitting unambiguous terms we call this decomposition \emph{monosemous PID from PED}.
Note that while this provides a decomposition of the joint mutual information into its redundant, unique and synergistic components (Eq.~\ref{eq:mi12pid}), it does not satisfy Eq.~\ref{eq:mi1pid}, that is the univariate mutual information cannot in general be obtained from the sum of the unique and redundant information.
This is because the univariate mutual informations includes partial entropy terms that do not appear in the joint mutual information\footnote{These ambiguous terms are: $H_\partial(\{1\}\{23\})$, $H_\partial(\{2\}\{13\})$, $H_\partial(\{12\}\{13\})$, $H_\partial(\{12\}\{23\})$, $H_\partial(\{12\}\{13\}\{23\})$.}.
Therefore, although this approach does not preserve one of the main properties of the PID (Eq.~\ref{eq:mi1pid}), we suggest it is nevertheless interesting and potentially useful.
First, the entropy decomposition perspective shines some light on the origin of some of the difficulties encountered when trying to obtain consistent PIDs.
Second, as we will show in the examples, this alternative approach has some interesting properties.
For example it provides qualitatively different decompositions to any existing measures for simple binary operations. 
These decompositions seem to give a better intuitive match to the functional differences in the examples (Section \ref{sec:binary}). 

\subsection{Identity axiom for PED based PIDs}
\label{sec:identity}

The canonical example to illustrate the conceptual problem with the original $I_\text{min}$ measure presented in \parencite{williams_nonnegative_2010} is the ``two-bit copy problem`` \parencite{harder_bivariate_2013,timme_synergy_2013,griffith_quantifying_2014}.
Consider two independent uniform binary variables $X_1$ and $X_2$ and define $S$ as a direct copy of these two variables $S=(X_1,X_2)$.
Since $X_1$ and $X_2$ are independent by construction there should be no overlapping redundant information.
This example led to the proposal of the identity axiom \parencite{harder_bivariate_2013} (Eq.~\ref{eq:idaxiom}).

As previously noted the entropy decomposition perspective suggests a modification to the axiom, to include only the redundant entropy part of mutual information and not the synergistic entropy (Eq.~\ref{eq:modidaxiom}).
Note that in the case of the original two-bit copy problem $I(X_1;X_2)=0$ and the two formulations are equivalent, since there can be no local misinformation terms\footnote{$I(X_1;X_2)=0 \implies P(x_1,x_2)=P(x_1)P(x_2) \forall x_1 \in X_1,x_2 \in X_2$} and hence no synergistic entropy.

In the case where $I(X_1;X_2)\neq0$ we can see that for $X_3=(X_1,X_2)$:
\begin{equation}
    H_\partial(\{1\}\{2\}\{3\}) = H_\cap(\{1\}\{2\}\{3\}) = H_\cap(\{1\}\{2\})
\end{equation}
and
\begin{equation}
    H_\partial(\{13\}\{23\}) = H_\partial(\{12\}\{13\}\{23\}) = 0
\end{equation}
since 
\begin{equation}
    H_\cap(\{3\}\{12\}) = H_\cap(\{12\}\{12\}) = H_\cap(\{12\}) = H(\{12\})
\end{equation}
so all nodes above $\{3\}\{12\}$ on the lattice must have zero partial entropy values.

Therefore, both the full (Eq.~\ref{eq:pedpid1}) and the monosemous (Eq.~\ref{eq:pedpid2}) PED based PIDs satisfy the modified identity axiom of Equation~\ref{eq:modidaxiom}.

We can also see that the PED derived measures satisfy symmetry in the choice of target for some terms \parencite{bertschinger_shared_2013}.
For the unique information terms $I_\partial(i; \{j\}) = I_\partial(j; \{i\})$. 
This is not obviously the case for the redundant information terms, but we have yet to find a counterexample.
It may be that there are additional constraints on the higher order terms of the lattice that could ensure target symmetry also for the redundancy term. 
The \textsc{and} example (Sec.~\ref{sec:and}) provides a counter example for target symmetry of the monosemous PED synergy term, but not for the full PED (where $I_\partial(3, \{12\})=I_\partial(1;\{23\})$).

\subsection{Quantifying mechanistic vs source redundancy}
\label{sec:mechanistic}

\textcite{harder_bivariate_2013} introduce a distinction between \emph{mechanistic} information redundancy, which arises from the mechanism of the considered system and \emph{source} or \emph{predictor} information redundancy, which arises from the relationship between the inputs and is independent of the mechanism. 

Source information redundancy is based on the idea that if $X_1$ and $X_2$ are dependent, this can contribute to the information redundancy measured between them about the target $X_3$.
For example, if $X_3 = X_1$ but $X_1$ and $X_2$ are correlated there will be information redundancy about $X_3$ between $X_1$ and $X_2$, but this arises solely from the relationship between $X_1$ and $X_2$ since the functional mechanism is independent of $X_2$. 
In contrast, in the case of a binary \textsc{and} gate, the inputs are independent, but it is generally accepted that there is information redundancy in this system (see Section \ref{sec:and}). 
This redundancy is therefore purely mechanistic, since it is induced by the functional properties of the system under study.

While the importance of this distinction has been recognised \parencite{harder_bivariate_2013,banerjee_synergy_2015,wibral_partial_2016} there are no existing proposals for how to define or quantify these different types of redundancy. 
We show here how mechanistic and source information redundancy can be explicitly separated within the PED based information decompositions. 

Similar to  Eq.~\ref{eq:mi1}:
\begin{equation}
    I(X_1;X_2) = H^{12}_\partial(\{1\}\{2\}) - H^{12}_\partial(\{12\}) \\
\label{eq:mix1x2}
\end{equation}
However, since we are here focussing on information redundancy between $X_1$ and $X_2$, following the same argument as for the identity axiom, we are concerned only with $H^{12}_\partial(\{1\}\{2\})$, since synergistic entropy effects between $X_1$ and $X_2$ should not affect the information redundancy between them.
From Figure~\ref{fig:ped2v3}:
\begin{equation}
    H^{12}_\partial(\{1\}\{2\}) = H_\partial(\{1\}\{2\}\{3\}) + H_\partial(\{1\}\{2\})
    \label{eq:mechent}
\end{equation}
where the terms on the right hand side are from the full three variable entropy decomposition.

Mechanistic information redundancy means there is more entropy shared between the three variables in the context of the full system, than there is between the inputs when the output is ignored. 
In partial entropy terms this corresponds to
\begin{equation}
    H_\partial(\{1\}\{2\}\{3\}) > H^{12}_\partial(\{1\}\{2\}) \implies  H_\partial(\{1\}\{2\}) < 0 
\end{equation}
where the implication follows from Eq.~\ref{eq:mechent}.
So mechanistic information redundancy corresponds to a negative $H_\partial(\{1\}\{2\})$ partial entropy term.
We hope this reasoning further illustrates why the monotonicity axiom, while at first glance appealing, is not appropriate for an entropy redundancy measure\footnote{We also suggest that it is not appropriate either for an information redundancy measure (see Section~\ref{sec:and}).}.
Any monotonic entropy redundancy measure would be by definition unable to detect mechanistic redundancy.

We therefore define 
\begin{align}
\begin{split}
    H^{\text{mech-}3}_\partial(\{1\}\{2\}\{3\}) &= \abs{ \min \left[ H_\partial(\{1\}\{2\}), 0 \right] } \\
    H^{\text{source-}3}_\partial(\{1\}\{2\}\{3\}) &= H_\partial(\{1\}\{2\}\{3\}) - H^{\text{mech-}3}_\partial(\{1\}\{2\}\{3\})
\end{split}
\end{align}
Note that since $H_\partial^{12}(\{1\}\{2\}) = H_\cap^{12}(\{1\}\{2\}) \geq 0$ both of the above terms are non-negative from Eq.~\ref{eq:mechent}.

Other partial entropy terms in $I_\partial(3; \{1\}\{2\})$ for both the full PED and mono-PED approaches all include synergistic entropy pairs with $X_3$.
These nodes lie above $\{1\}\{2\}$ on the entropy lattice, and so they cannot quantify source redundancy. 
We therefore assign these terms to the mechanistic part of the redundant partial information.
For both approaches the source information redundancy is the same:
\begin{equation}
    I_\partial^\text{source}(3; \{1\}\{2\}) = H^{\text{source-}3}_\partial(\{1\}\{2\}\{3\})
\end{equation}
while the mechanistic information redundancy is:
\begin{align}
    I_\partial^\text{mech}(3; \{1\}\{2\}) &=  H^{\text{mech-}3}_\partial(\{1\}\{2\}\{3\}) - H_\partial(\{13\}\{23\}) - H_\partial(\{12\}\{13\}\{23\}) \\
    I_\partial^\text{mech}(3; \{1\}\{2\}) &=  H^{\text{mech-}3}_\partial(\{1\}\{2\}\{3\}) - H_\partial(\{13\}\{23\})
\end{align}
for the full and monosemous PED approaches respectively.

\begin{figure}[htbp]
    \centering
    \includegraphics[width=0.8\textwidth]{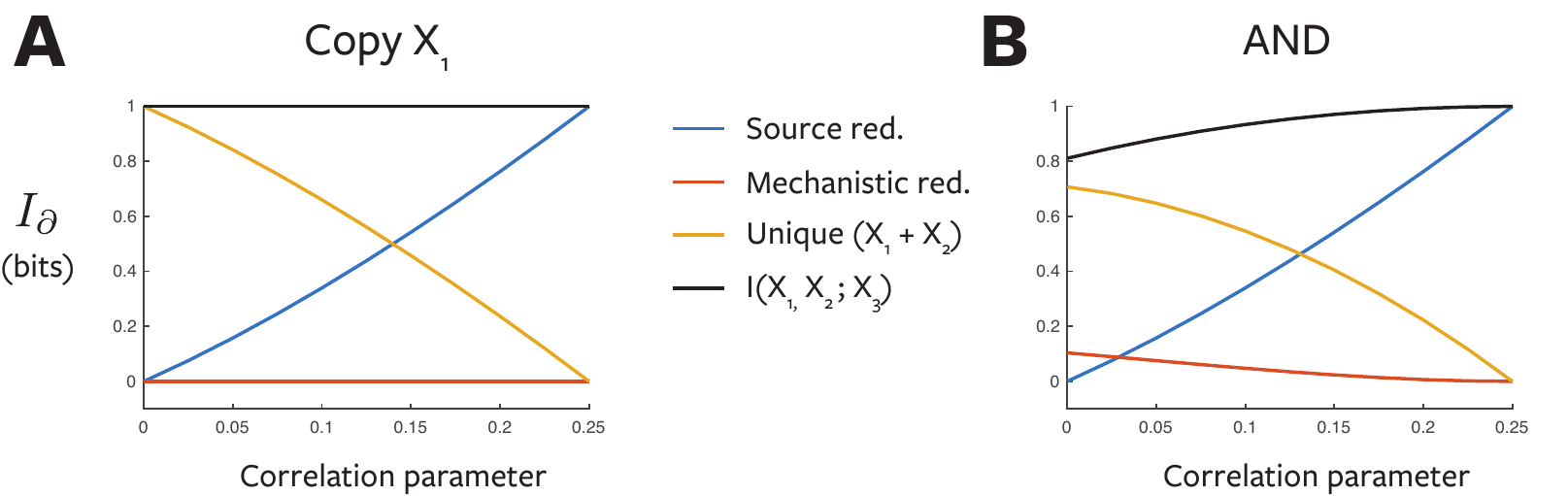}
    \caption{\emph{Mechanistic vs source redundancy from monosemous PED partial information.} Non-zero partial entropy terms are shown for two example systems as a function of the dependence between the inputs (Eq.~\ref{eq:mechsystems}). (\textbf{A}) Output $X_3$ = $X_1$. (\textbf{B}) Output $X_3$ is $X_1$ \textsc{and} $X_2$. With the monosemous-PED neither system shows any synergy. For copy (A), $I_\partial(3;\{2\})=0$. For \textsc{and} (B) $I_\partial(3;\{1\}) = I_\partial(3;\{2\})$ and the sum of the two terms is plotted.} 
    \label{fig:mechvssource}
\end{figure}

Figure \ref{fig:mechvssource} shows these values for the two example systems introduced above.
The first system the output is a copy of $X_1$, the second system is an \textsc{and} gate. 
In both cases, the relationship between the two inputs is modulated while preserving uniform marginals by using joint distributions of the form:
\begin{align}
\begin{split}
    P(X_1=0,X_2=0) &= P(X_1=1,X_2=1) = 0.25 + c \\
    P(X_1=0,X_2=1) &= P(X_1=1,X_2=0) = 0.25 - c
\end{split}
\label{eq:mechsystems}
\end{align}
where $c$ is a correlation parameter in the range $[0, 0.25]$.
When $c=0$ the two inputs are independent, when $c=0.25$, $X_2$ is a copy of $X_1$.
In the copy system, there is no mechanistic redundancy, and the joint mutual information transitions from pure unique information to pure source redundancy as the dependence between the inputs is increased. 
Similarly for \textsc{and} the source redundancy increases with dependence, while mechanistic redundancy (present with independent inputs) and unique information decrease.

\subsection{Redundant entropy as a measure of dependence: Pure mutual information}
\label{sec:puremi}

We have seen that mutual information is the difference between redundant and synergistic entropy (Eq.~\ref{eq:mi_ped}).
However, we contend that mutual information is usually interpreted as redundant entropy.
For example, when first coining the term \emph{mutual information}, \textcite{fano_statistical_1959} states: ``Clearly it is a measure of the extent to which the two events are more likely to occur together than if they were statistically independent''.
However, technically this interpretation applies only to redundant entropy.
Mutual information subtracts the synergistic entropy, a measure of the extent to which two events are less likely to occur together.
Bearing in mind surprisal was originally termed \emph{self-information}, the actual term ``mutual information'' is almost a synonym for ``shared entropy''.
We suggest that many of the difficulties around the PID, including the focus on non-negativity and the lack of clarity over basic properties that an information redundancy measure should satisfy, come from this long-standing misinterpretation of mutual information as shared entropy, without a full appreciation that it also quantifies synergistic effects.
We hope the PED perspective presented here can help clarify this long held incorrect interpretation of mutual information.

Exactly as interaction information conflates redundant information and synergistic information, and so is unsatisfactory as a measure of shared information \parencite{williams_nonnegative_2010}, we propose here that mutual information, since it conflates redundant and synergistic entropy is not a pure measure of shared entropy.
While redundant entropy (like redundant information) is not accessible with linear combinations of classical Shannon quantities \parencite{james_multivariate_2016}, the Partial Entropy Decomposition provides a way to quantify it.
We therefore propose that the redundant entropy term itself can be used as a measure of bivariate dependence, that truly measures only shared, and not synergistic effects.

While this is simply redundant entropy, in this context we term it \emph{pure mutual information} in order to preserve familiar notation related to mutual information:

\newcommand{\IredH}{I^{p}}
\begin{equation}
    \IredH(X_1;X_2) = H^{12}_\partial(\{1\}\{2\})
\end{equation}

Pure mutual information satisfies many of the fundamental properties of mutual information.
\begin{align}
\IredH(X; Y) &\geq 0 \\
\IredH(X; Y)=0 &\iff p(x,y) = p(x)p(y) \forall x,y
\end{align}
and with conditional pure mutual information defined as
\begin{align}
\begin{split}
    \IredH(X;Y|Z) &= \IredH(X; Y,Z) - \IredH(X; Z) \\
    &=  H^{XYZ}_\partial(\{X\}\{Y\}) + H^{XYZ}_\partial(\{X\} \{YZ\})
\end{split}
\end{align}
it also satisfies the chain rule \parencite{cover_elements_1991}.

\begin{figure}[htbp]
    \centering
    \includegraphics[width=0.8\textwidth]{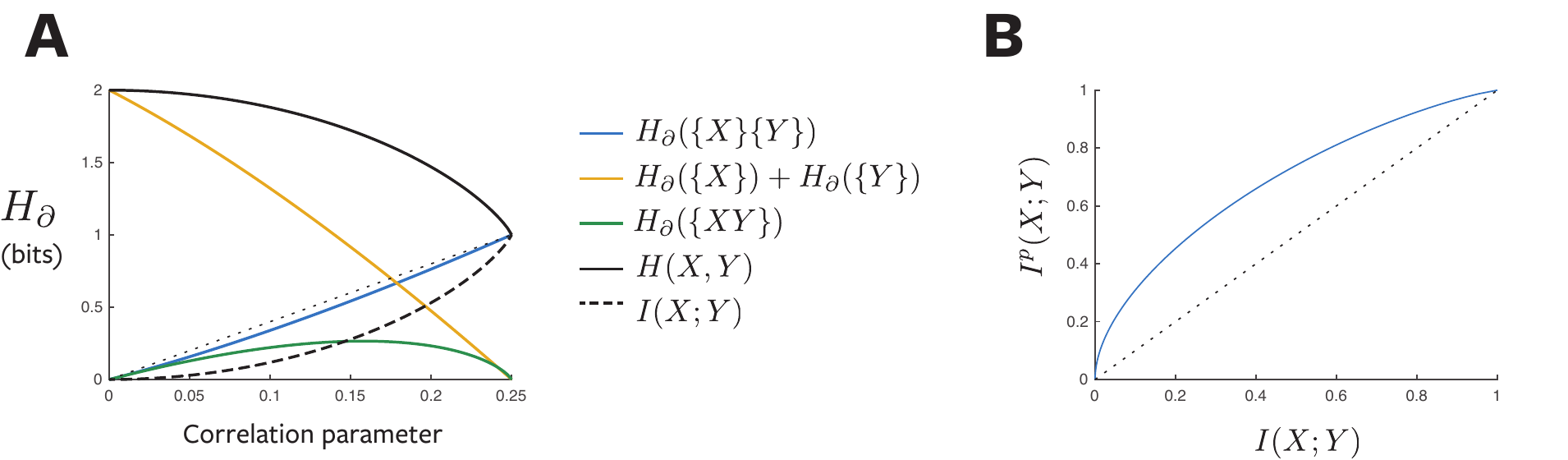}
    \caption{\emph{Partial Entropy Decomposition for two binary variables.}.} 
    \label{fig:puremibinary}
\end{figure}

To illustrate this we consider a bivariate binary system with uniform marginals, where the relationship between the two variables is parametrically modulated following Eq.~\ref{eq:mechsystems}.
Figure \ref{fig:puremibinary}A shows the full partial entropy decomposition for this system as a function of the correlation probability parameter $c$. 
As the variables become more dependent, total entropy and unique partial entropy decrease; redundant partial entropy and mutual information between the two increase. 
However, synergistic entropy is $0$ for both independent and fully dependent inputs, but rises for intermediate levels of dependence, peaking at $0.27$ bits when $c=0.16$ ($p(0,0)=p(1,1)=0.41$).
This gives the mutual information the curved form (the redundant entropy is close to linear).
Figure \ref{fig:puremibinary}B shows the relationship between mutual information and pure mutual information for this family of systems.
Pure mutual information is greater than and monotonically increasing as a function of mutual information.
We conjecture that these properties hold in general.

While we suggest pure mutual information provides an interesting perspective on mutual information as a measure of dependence between two systems, since the contrast with mutual information makes explicitly clear the effects of the inclusion of synergistic entropy, it is unlikely to satisfy many of the properties and theorems relating mutual information to coding over noisy channels.
In this perspective, synergy is likely to be important as it would limit the performance of possible coding strategies.

\subsubsection{Partial information decomposition of pure mutual information}

Without synergistic contributions, most of the natural intuitions that have been incorrectly applied to decomposition of mutual information, do apply to pure mutual information.
For example, the identity axiom is now correct for $\IredH$ as originally formulated:
\begin{equation}
    \IredH_\cap\left(\{\mathbf{A_1},\mathbf{A_2}\}; \mathbf{A_1},\mathbf{A_2}\right) = \IredH(\mathbf{A_1}; \mathbf{A_2})
    \label{eq:idaxiompure}
\end{equation}

Further, since
\begin{equation}
    \IredH(X_1,X_2 ; X_3) = H_\partial(\{1\}\{2\}\{3\}) + H_\partial(\{1\}\{3\}) + H_\partial(\{2\}\{3\}) + H_\partial(\{12\}\{3\})
\end{equation}
a partial information decomposition is obtained trivially from the four constituent terms.
This satisfies both Eq.~\ref{eq:mi12pid} and Eq.~\ref{eq:mi1pid} for pure mutual information.

In all the examples presented here, the PID for pure mutual information is equal to the monosemous PED based PID, so we do not explicitly report the values, although this might not be true in general.
However, the purpose of introducing pure mutual information is to provide a perspective in which the results obtained from the monosemous PID do provide a decomposition satisfying Eq.~\ref{eq:mi1pid}, if you consider pure mutual information (redundant entropy) as the measure of dependence to be decomposed.

While there are obviously many aspects of mutual information that depend on its inclusion of synergistic effects, we hope that considering pure mutual information is a useful exercise to explicitly demonstrate that it is the synergistic entropy which makes obtaining a partial information decomposition challenging.

\section{Examples}

Code implementing the methods and all examples is available at:

\href{https://github.com/robince/partial-info-decomp}{\texttt{https://github.com/robince/partial-info-decomp}}.

To calculate $I_\text{broja}$ and compute the maximum entropy distributions under pairwise marginal constraints we use the \texttt{dit} package \parencite{james_dit/dit_2017}\footnote{\href{https://github.com/dit/dit}{\texttt{https://github.com/dit/dit}}
\hspace{0.2cm}  \href{http://docs.dit.io/}{\texttt{http://docs.dit.io/}}}.

\subsection{Dyadic vs Triadic structure}
\label{sec:dyadictriadic}

\begin{figure}[htbp]
    \centering
    \includegraphics[width=0.8\textwidth]{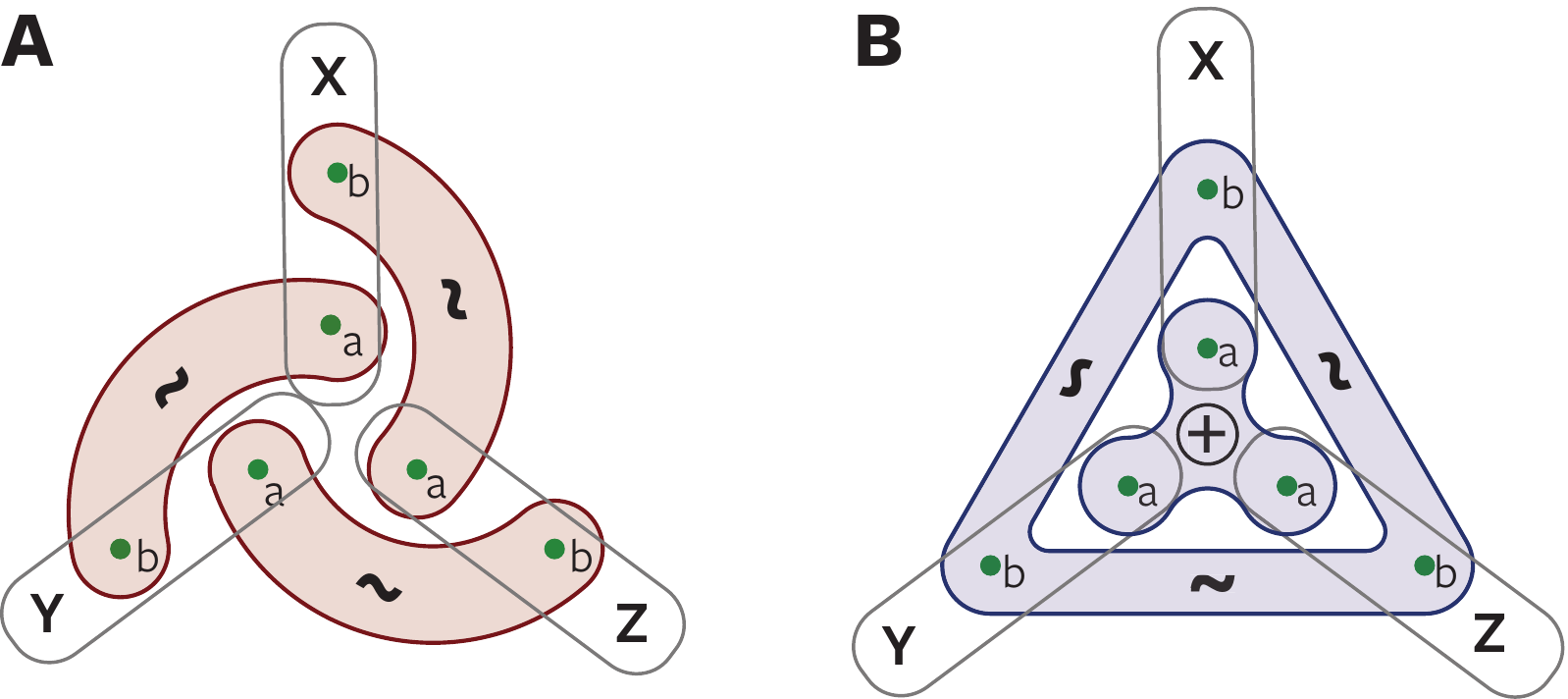}
    \caption{\emph{Distributions with equivalent classical information measures but dyadic (\textbf{A}) vs triadic (\textbf{B}) structure.} Each variable $X$, $Y$ and $Z$ consists of two bits, labelled $a$ and $b$. $\sim$ indicates bits which are coupled (distributed identically) and $\oplus$ indicates the enclosed variables form the \textsc{xor} relation.  Modified from \parencite{james_multivariate_2016}.}
    \label{fig:dyadictriadic}
\end{figure}

We first consider two example systems which have the same classical information diagrams, but with a dyadic vs triadic structure \parencite{james_multivariate_2016}.
The challenge posed by \textcite{james_multivariate_2016} is to decompose the entropy of these multivariate systems in a way that meaningfully represents their different generative structures.   
While the PID falls short of this requirement because of its asymmetry (and it can only account for the two bits of mutual information, and not the third bit of entropy also present in the system) we believe the partial entropy decomposition presented here directly addresses this question. 
The two systems are illustrated in Figure~\ref{fig:dyadictriadic} \parencite{james_multivariate_2016}.
Both consist of three variables ($X$, $Y$, $Z$) each consisting of two uniform bits ($a$, $b$). 
In the dyadic system (Fig.~\ref{fig:dyadictriadic}A) one bit is shared between each pair of variables.
In the triadic system (Fig.~\ref{fig:dyadictriadic}B) one bit is shared betwen all three variables, and the other bit forms an \textsc{xor} relationship between the three.  

\begin{figure}[htbp]
    \centering
    \includegraphics[width=0.5\textwidth]{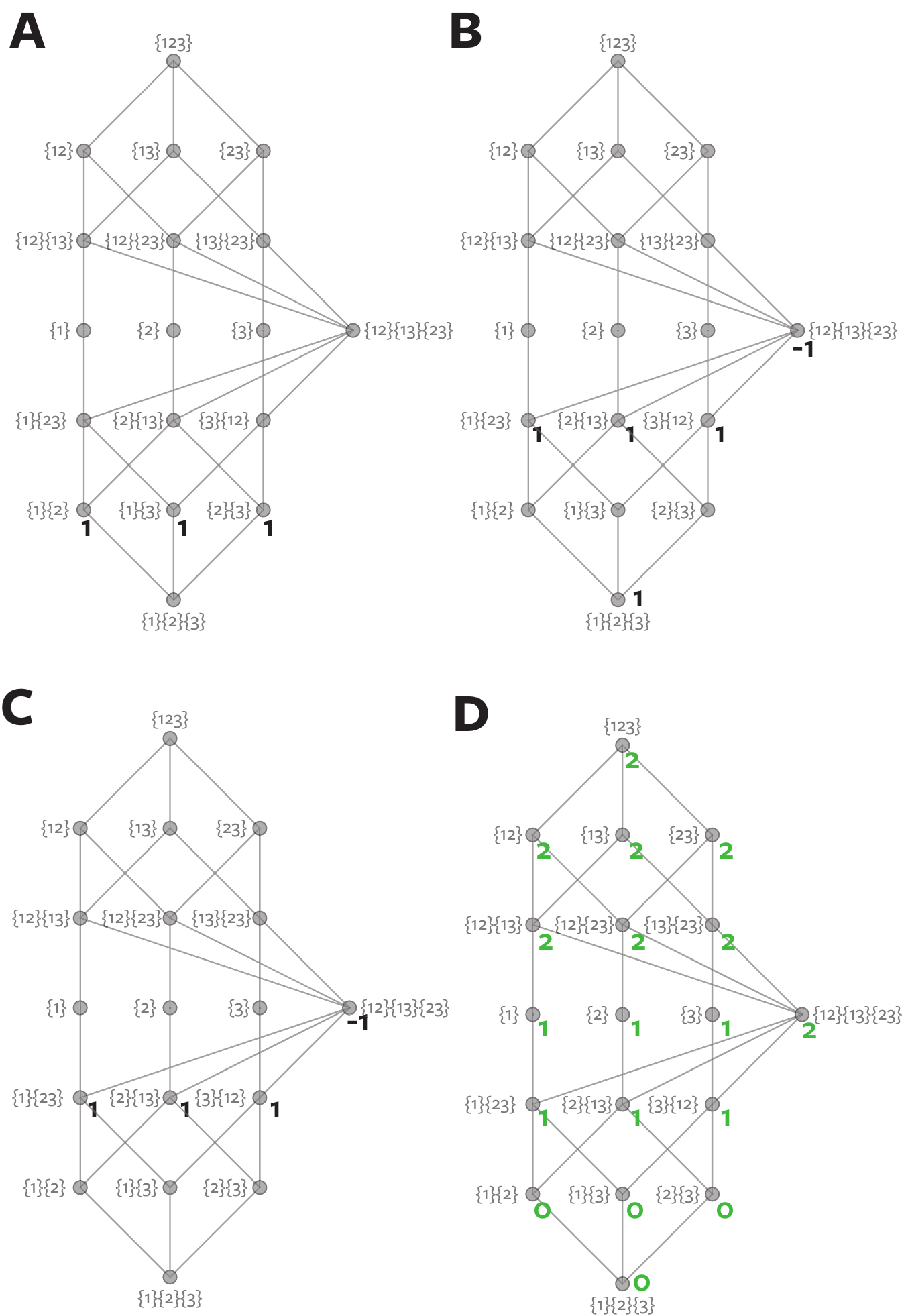}
    \caption{\emph{Partial Entropy Decomposition for (\textbf{A}) dyadic system (Fig.~\ref{fig:dyadictriadic}A), (\textbf{B}) triadic system  (Fig.~\ref{fig:dyadictriadic}B), (\textbf{C}) \textsc{xor}.} Nodes are labelled with their partial entropy values in bits. Nodes with zero values are not labelled. (\textbf{D}) $H_\text{cs}$ values for the \textsc{xor} system.}
    \label{fig:dyadictriadicped}
\end{figure}

The results of the partial entropy decomposition for these systems, and for a related pure \textsc{xor} system (3 single bit uniform variables in an \textsc{xor} configuration) are shown in Figure \ref{fig:dyadictriadicped}.
We can see the PED reveals the different structure of the distributions in a way that matches their generative structure.
In the dyadic example (Fig. \ref{fig:dyadictriadicped}A), 1 bit is allocated to each pairwise univariate redundancy term, reflecting the three pairwise coupled bits which form the system.
In the triadic example (Fig. \ref{fig:dyadictriadicped}B), at first glance the decomposition is a little harder to interpret, due to the presence of the negative term for the entropy commonly available in all three synergistic pairs of variables.
However, from comparison with the decomposition for a pure 3 variable binary \textsc{xor} (Fig. \ref{fig:dyadictriadicped}C) we can see the triadic decomposition correctly reflects the single bit shared between all variables, together with 2 bits from an \textsc{xor} structure.
We therefore suggest this entropy decomposition approach addresses the challenge posed by \textcite{james_multivariate_2016} of how to meaningfully describe the multivariate structure of these two systems.

Note that here the use of pairwise maximum entropy distribution is necessary to obtain the single triple-wise redundant bit in the triadic system.

The \textsc{xor} decomposition structure consists of 1 bit available synergistically from each pair of variables that is redundant with the third variable. 
However, since these three nodes add up to $3$ bits, when there are in total only $2$ bits of entropy for \textsc{xor}, there is a negative term for the three-way overlap of pairwise synergies (Fig. \ref{fig:dyadictriadicped}C).
Note that although $H_\text{cs}$ is not in general monotonic (see example \textsc{and} below), in the case of \textsc{xor} the values are monotonic on the lattice (Fig. \ref{fig:dyadictriadicped}D), so non-monotonicity does not cause the negative term in this case.
At first glance this is hard to interpret; when a similar situation arose in the information decomposition with $I_\text{ccs}$ \parencite{ince_measuring_2016} we suggested a normalisation procedure to obtain a non-negative decomposition. 
However, the $H_\text{cs}$ values on the lattice (Fig. \ref{fig:dyadictriadicped}D)  match the properties of the \textsc{xor}: 1 bit is available in each variable alone, and the system is fully determined (2 bits) by any synergistic pair. 
So the positive partial entropy values in the PED (Fig. \ref{fig:dyadictriadicped}C) seem correct --- it is true that each of the positive nodes do have 1 bit of entropy which is not available from any lower nodes on the lattice. 
$H_\partial(\{12\}\{13\}\{23\})=-1$ indicates that the redundant entropy shared between the three synergistic pairs is actually less than that which is expected from the lower terms on the lattice.
It corrects the double counting of the two \textsc{xor} bits in the lower level level of the lattice.
In Section \ref{sec:mechanistic} we noted how negative values for $H_\partial(\{1\}\{2\})$ correspond to mechanistic redundancy.
While we haven't fully developed the relationship we suggest that this negative value can also be interpreted as a mechanistic redundancy effect, at the level of pairwise synergistic entropy.

The full three variable lattice shown here contains 18 values. 
While the full decomposition can reveal any asymmetries in the multivariate system we suggest that in many cases it might be useful to consider a simplified presentation based on the order structure \parencite{ince_measuring_2016}, that is considering the sum over all terms of a specific order. This reduces the three-variable entropy decomposition to 8 terms:
\begin{equation}
    \begin{alignedat}{9}
     &[ (1,1,1) &&: (1,1) &&: (1,2) &&: (1) &&: (2,2,2) &&: (2,2)  &&: (2)  &&:  (3) &&] \\
     \textsc{dyadic} = &[ 0 &&: 3 &&: 0 &&: 0 &&: 0 &&: 0 &&: 0 &&: 0  &&] \\
     \textsc{triadic} = &[ 1 &&: 0 &&: 3 &&: 0 &&: -1 &&: 0 &&: 0 &&: 0  && ] \\
     \textsc{xor} = &[ 0  &&: 0 &&: 3 &&: 0 &&: -1 &&: 0 &&: 0 &&: 0  &&] \\
\end{alignedat}
\label{eq:ditriorder}
\end{equation}

\subsection{Binary logical operations}

\label{sec:binary}

\subsubsection{XOR}
\label{sec:xor}

We start with \textsc{xor}, which has the PED shown in Figure \ref{fig:dyadictriadicped}C.
Table~\ref{tab:pidxor} shows the PIDs for this system.

\begin{table}[htbp]
    \centering
    \begin{tabular}{| c  | c | c | c | c | c |}
        \hline
        \textbf{Node} & $\bm{I_\partial} [ I_\text{min} ]$ & $\bm{I_\partial} [ I_\text{broja} ]$ & $\bm{I_\partial} [ I_\text{ccs} ]$ & $\bm{I_\partial} [ \textsc{ped} ]$ & $\bm{I_\partial} [ \text{\footnotesize mono-}\textsc{ped} ]$ \\
        \hline \hline
        $\{12\}$ &     $1$ & $1$ & $1$ & $2 $ & $1$ \\ 
        $\{1\}$ &      $0$ & $0$ & $0$ & $-1$ & $0$ \\
        $\{2\}$ &      $0$ & $0$ & $0$ & $-1$ & $0$ \\ 
        $\{1\}\{2\}$ & $0$ & $0$ & $0$ & $1$  $(0,1)$ & $0$ $(0,0)$ \\ \hline
     \end{tabular}
     \caption{\emph{PIDs for \textsc{xor}.} Bracketed redundancy values decompose the redundant partial information into source and mechanistic redundancy respectively.}
  \label{tab:pidxor}
\end{table}

Note that in this case both the full and the monosemous PED (denoted mono-\textsc{ped}) based decompositions satisfy Eq.~\ref{eq:mi1pid}, as well as Eq.~\ref{eq:mi12pid}.
However while mono-PED matches the common decomposition obtained with the other methods, the full PID from PED is strikingly different.
At first glance the full PED PID seems counter intuitive based on the properties of \textsc{xor}.
The $1$ bit of unique misinformation in $X_1$ comes from the term $H_\partial(\{2\}\{13\})$. 
This is an entropy term that has a value of $1$ bit, represents misinformation between $X_1$ and $X_3$, and appears in $I(X_1;X_3)$ (but not in $I(X_2;X_3)$). 
Therefore, in this entropy perspective there is indeed $1$ bit of unique misinformation in each variable.
The single redundant bit also seems counter-intuitive since in the predictor-target perspective $X_1$ and $X_2$ are independent. 
It arises from the negative three way redundant pairwise synergy term, which represents a mechanistic redundancy between pairwise synergistic entropies.
Indeed, in the source vs mechanistic split it appears as a mechanistic redundancy.
The $2$ bits of synergy arise from the sum of all non-zero terms in the lattice, which appear in Eq.~\ref{eq:pedpid1syn} with positive sign. 
Note that the monosemous PED approach gives in a sense the net-effects, here there is overall just $1$ bit of synergy, but the full description afforded by the resolution of the partial entropy approach shows that even in this simple example system there is both redundancy and synergy at the level of the individual target-predictor mutual informations.

\subsubsection{AND}
\label{sec:and}

\begin{figure}[htbp]
    \centering
    \includegraphics[width=0.5\textwidth]{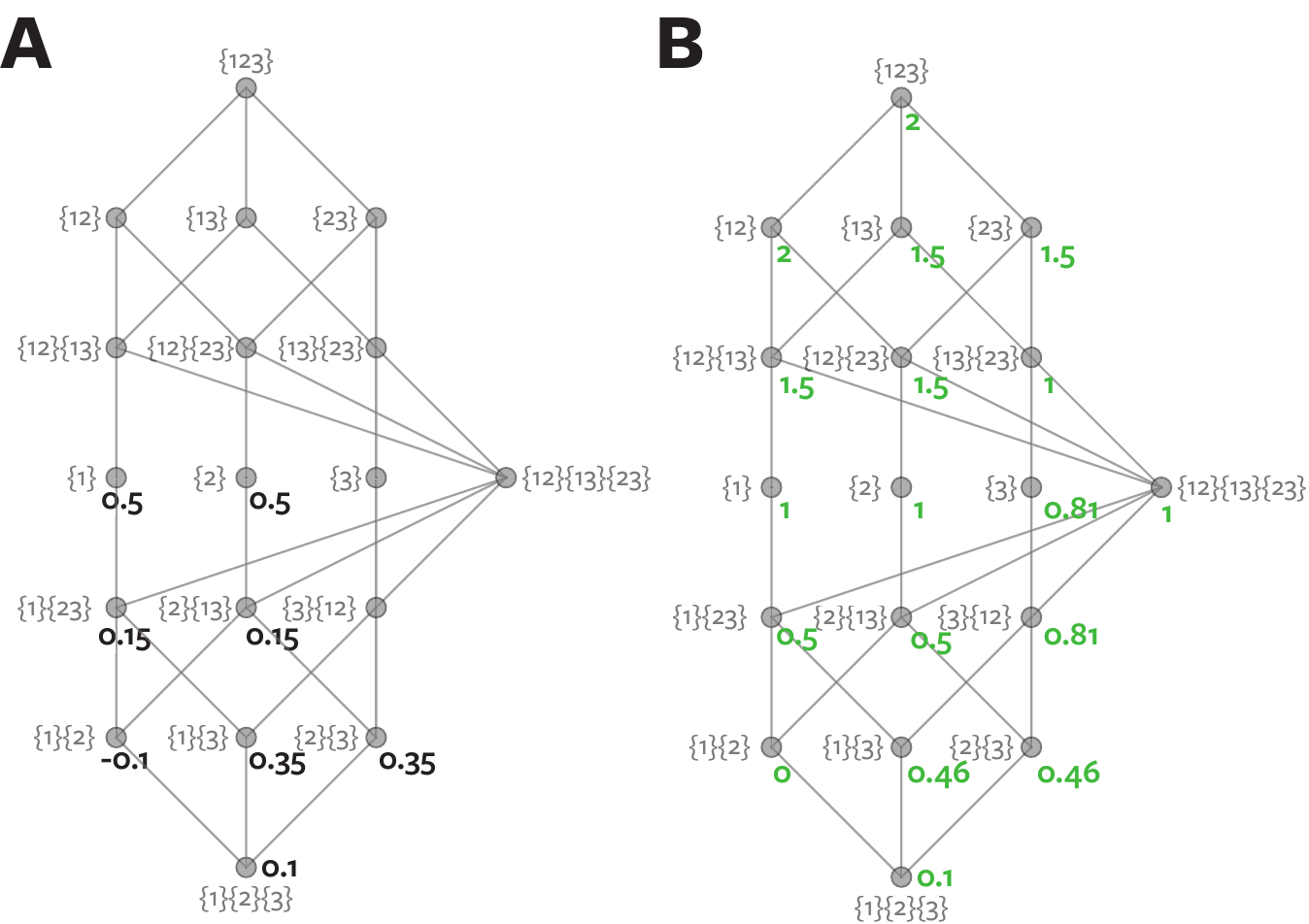}
    \caption{\emph{Partial Entropy Decomposition for \textsc{and}.} (\textbf{A}) Nodes are labelled with their partial entropy values in bits. Nodes with zero values are not labelled. (\textbf{B}) $H_\text{cs}$ values for each node are labelled in green.}
    \label{fig:andped}
\end{figure}

We next consider \textsc{and}, which is equivalent in information theoretic terms to \textsc{or} \parencite{ince_measuring_2016}.
Figure~\ref{fig:andped} shows the PED and the values of $H_\text{cs}$ on the entropy lattice.
The PED reflects the asymmetry of the system. 
Note that $H_\partial(\{1\}\{2\})$ is negative, and this results from non-monotonic $H_\text{cs}$ values on the lattice. 
$H_\text{cs}(\{1\}\{2\}) = 0 < H_\text{cs}(\{1\}\{2\}\{3\})=0.1$.
Note that these values are not affected by the use of the pairwise maximum entropy distribution, the same non-monotonicity occurs if the full distribution is used for the three way redundancy node.
The three way univariate redundancy arises from a single non-zero pointwise term when $X_1 = X_2 = X_3 = 0$.
In the entropy perspective, there is redundant local entropy in the context of the three variable system, which is removed when variable $X_3$ is marginalised away in the calculation of $H_\cap(\{1\}\{2\})$: therefore representing mechanistic information redundancy. 

Table~\ref{tab:pidand} shows PIDs for this system. 
Note here that all the pointwise approaches allocate non-zero unique information to both variables, in contradiction of Corollary 8 in \textcite{bertschinger_quantifying_2014}.
However, we suggest that equality of target-predictor marginals only ensures that the two unique information values should be equal, not that they must be zero.
Here neither of the PED based approaches agree with $I_\text{ccs}$.
However, the full PED decomposition is equivalent to that obtained with $I_\text{ccs}$ if the full distribution is used for the redundancy term, rather than $P_\text{ind}$ \parencite{ince_measuring_2016}.
In this case, the discrepancy arises because of the asymmetry in the system, and the fact that for this system, $P_\text{ind}$ is not equal to the second order marginal preserving maximum entropy distribution.
Both PED approaches correctly identify the redundancy as purely mechanistic; it must be mechanistic since the inputs are independent.
Another interesting property is that the monosemous PED decomposition allocates zero synergy.
Comparing the values for this decomposition for \textsc{and} and \textsc{xor} suggests that this approach can qualitatively discriminate \textsc{and} and \textsc{xor} based on the synergy term alone.
Note that for the mono-\textsc{ped} here, the sum of the unique and redundant partial information terms does not equal the univariate mutual information (Eq.~\ref{eq:mi1pid}).
Therefore the full interpretation of the PID for mutual information can not be applied (although it does apply for pure mutual information, Section \ref{sec:puremi}). 
Nevertheless, it is a valid and in this case non-negative decomposition of the joint mutual information into its unambiguous redundant, unique and synergistic contributions.

\begin{table}[htbp]
    \centering
    \begin{tabular}{| c | c | c | c | c | c |}
        \hline
        \textbf{Node} & $\bm{I_\partial} [ I_\text{min} ]$ & $\bm{I_\partial} [ I_\text{broja} ]$ & $\bm{I_\partial} [ I_\text{ccs} ]$ & $\bm{I_\partial} [ \textsc{ped} ]$ & $\bm{I_\partial} [ \text{\footnotesize mono-}\textsc{ped} ]$ \\
        \hline \hline
        $\{12\}$ &     $0.5$ & $0.5$ & $0.43$ & $0.29$ & $0$ \\ 
        $\{1\}$ &      $0$ & $0$ & $0.07$ & $0.21$ & $0.35$ \\
        $\{2\}$ &      $0$ & $0$ & $0.07$ & $0.21$ & $0.35$ \\ 
        $\{1\}\{2\}$ & $0.31$ & $0.31$ & $0.24$ & $0.10$ $(0,0.1)$ & $0.10$ $(0,0.1)$ \\ \hline
     \end{tabular}
     \caption{\emph{PIDs for \textsc{and}/\textsc{or}.} Bracketed redundancy values decompose the redundant partial information into source and mechanistic redundancy respectively.}
  \label{tab:pidand}
\end{table}

We now consider the same \textsc{and} distribution, but we switch the target variable to be one of the two uniform independently distributed input bits. 
PIDs for this system are shown in Table~\ref{tab:pidandswitch}.
We denote here the two inputs to the \textsc{and} as $X$, $Y$, and the output as $Z$.
Now all three pre-existing methods considered agree, allocating $0.31$ bits of unique information about $Y$ to $Z$ together with $0.19$ bits of synergy about $Y$ between $X$ and $Z$.
This shows the asymmetry of PID in terms of the selected target variable \parencite{james_multivariate_2016}.
The unique information in $Y$ about target $Z$ (0 bits) is not equal to the unique information in $Z$ about target $Y$ (0.31 bits).
Similarly, the redundancy between $X$ and $Y$ about $Z$ (0.31 bits) is not equal to the redundancy between $X$ and $Z$ about $Y$ (0 bits).
The PED approach removes these asymmetries, since the entropy lattice is preserved by reordering the variables (the terms change location as the variables are permuted but the values remain the same).

However, now the unique information in $X$ about $Y$ is negative ($-0.1$ bits) from the $H_\text{cs}(\{1\}\{2\})$ term in Figure~\ref{fig:andped}.
This suggests there is effective misinformation between the marginally independent inputs $X$ and $Y$ in the context of the full three variable system.
In fact, when considering $P_\text{ind}$, the maximum entropy distribution preserving the $X,Z$ and $Y,Z$ pairwise marginals \parencite{ince_measuring_2016}, there is a dependence indicated by non-zero mutual information between $X$ and $Y$ under that distribution, and this mutual information calculation includes pointwise negative misinformation terms.
For the reasons discussed above, the asymmetric $P_\text{ind}$ gives a different perspective to the PED used here, so the terms are not equal numerically, but this at least suggests the potential for misinformation between $X$ and $Y$ in the context of the full system.

The same amount of redundancy is present as for \textsc{and}.
However, the redundancy here is correctly identified as source redundancy not mechanistic redundancy, since it arises from the dependence between $X$ and $Z$.

\begin{table}[htbp]
    \centering
    \begin{tabular}{| c | c | c | c | c | c |}
        \hline
        \textbf{Node} & $\bm{I_\partial} [ I_\text{min} ]$ & $\bm{I_\partial} [ I_\text{broja} ]$ & $\bm{I_\partial} [ I_\text{ccs} ]$ & $\bm{I_\partial} [ \textsc{ped} ]$ & $\bm{I_\partial} [ \text{\footnotesize mono-}\textsc{ped} ]$ \\
        \hline \hline
        $\{12\}$ &     $0.19$ & $0.19$ & $0.19$ & $0.29$ & $0.15$ \\ 
        $\{1\}$ &      $0$ & $0$ & $0$ & $-0.10$ & $-0.10$ \\
        $\{2\}$ &      $0.31$ & $0.31$ & $0.31$ & $0.21$ & $0.35$ \\ 
        $\{1\}\{2\}$ & $0$ & $0$ & $0$ & $0.10$ $(0.1,0)$ & $0.10$ $(0.1,0)$ \\ \hline
     \end{tabular}
     \caption{\emph{PIDs for \textsc{and}/\textsc{or} when predicting one of the inputs to the gate from the output and the other input.} Bracketed redundancy values decompose the redundant partial information into source and mechanistic redundancy respectively.}
  \label{tab:pidandswitch}
\end{table}

\subsubsection{SUM}

While not strictly a logic gate we consider \textsc{sum}, addition of two binary inputs to form a ternary output, as an extension of \textsc{and}. 
This binary summation operation is equivalent to the system termed \textsc{XorAnd} in \textcite{bertschinger_quantifying_2014}.
Table~\ref{tab:pidsum} shows PIDs for this system. 
In contrast to \textsc{xor} here it is the full PED which agrees with existing approaches (but not $I_\text{ccs}$), while the monosemous version gives a different perspective.
In this case, the monosemous PED approach does satisfy Eq.~\ref{eq:mi1pid} and shows there is no unambiguous redundancy between the two summands; each has $0.5$ bits of unique information and there is $0.5$ bits of information available synergistically.
The non-zero PED terms for this case are:
\begin{align}
\begin{split}
    H_\partial(\{1\}\{3\}) = H_\partial(\{2\}\{3\}) = H_\partial(\{1\}\{23\}) = H_\partial(\{2\}\{13\}) = H_\partial(\{3\}\{12\}) = 0.5 \\
    H_\partial(\{12\}\{13\}\{23\}) = -0.5
\end{split}
\label{eq:pedsum}
\end{align}

\begin{table}[htbp]
    \centering
    \begin{tabular}{| c | c | c | c | c | c |}
        \hline
        \textbf{Node} & $\bm{I_\partial} [ I_\text{min} ]$ & $\bm{I_\partial} [ I_\text{broja} ]$ & $\bm{I_\partial} [ I_\text{ccs} ]$ & $\bm{I_\partial} [ \textsc{ped} ]$ & $\bm{I_\partial} [ \text{\footnotesize mono-}\textsc{ped} ]$ \\
        \hline \hline
        $\{12\}$ &     $1$ & $1$ & $0.79$ & $1$ & $0.5$ \\ 
        $\{1\}$ &      $0$ & $0$ & $0.21$ & $0$ & $0.5$ \\
        $\{2\}$ &      $0$ & $0$ & $0.21$ & $0$ & $0.5$ \\ 
        $\{1\}\{2\}$ & $0.5$ & $0.5$ & $0.29$ & $0.5$ $(0,0.5)$ & $0$ $(0,0)$ \\ \hline
     \end{tabular}
     \caption{\emph{PIDs for \textsc{sum}.} Bracketed redundancy values decompose the redundant partial information into source and mechanistic redundancy respectively.}
  \label{tab:pidsum}
\end{table}

\subsubsection{Summary}
\label{sec:binsummary}
Together these examples show that the entropy decomposition perspective, particularly the PID based on the monosemous PED terms, opens a new window on the structure of redundancy and synergy in these simple systems. 
We suggest that the PID based on the monosemous PED terms may be most useful, as it seems to provide an intuitive decomposition of these systems (Table~\ref{tab:pidpedlogic}).
It shows \textsc{xor} has only synergistic information, \textsc{and} has no synergistic component in $I(X_1,X_2;X_3)$, and summation has no redundancy with equal amounts of unique information within and synergy between the summands.
This shows that the thresholding of \textsc{sum} to result in the logical \textsc{and} removes the synergy between the two inputs and introduces some mechanistic information redundancy.

\begin{table}[htbp]
    \centering
    \begin{tabular}{| c | c | c | c |}
        \hline
        \textbf{Node} & \textsc{xor} & \textsc{and/or} & \textsc{sum} \\
        \hline \hline
        $\{12\}$ &     $1$ & $0$ & $0.5$ \\ 
        $\{1\}$ &      $0$ & $0.35$ & $0.5$ \\
        $\{2\}$ &      $0$ & $0.35$ & $0.5$ \\ 
        $\{1\}\{2\}$ & $0$ & $0.10$ (mechanistic) & $0$ \\ \hline
     \end{tabular}
     \caption{\emph{mono-\textsc{PED} based PID for \textsc{xor}, \textsc{and} and \textsc{sum}.}}
  \label{tab:pidpedlogic}
\end{table}

\subsection{Williams and Beer (2010) Examples}

\begin{figure}[htbp]
    \centering
    \includegraphics[width=0.5\textwidth]{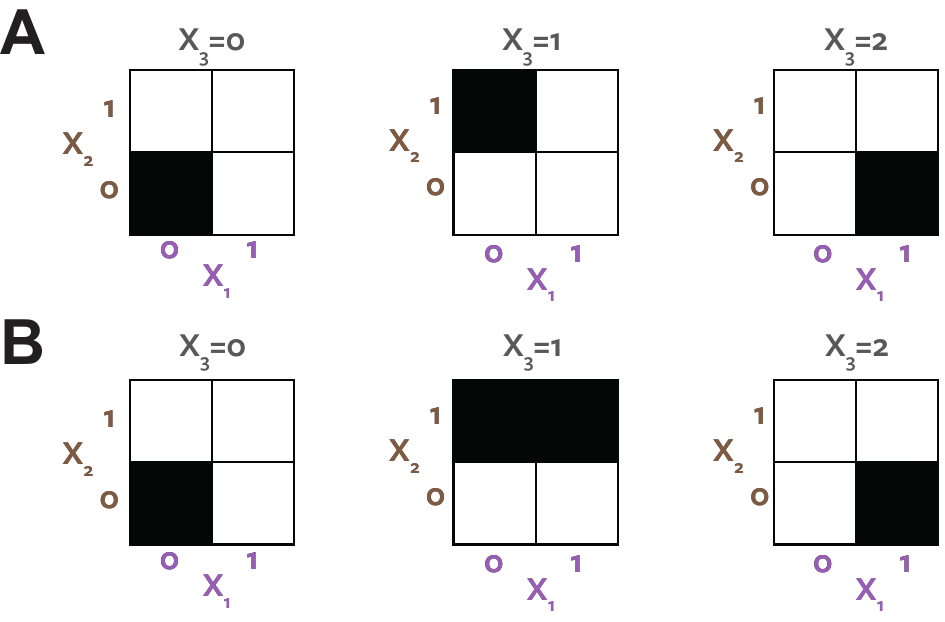}
    \caption{\emph{Probability distributions for two example systems}. 
    Black tiles represent equiprobable outcomes.
    White tiles are zero-probability outcomes.
    Modified from \textcite{williams_nonnegative_2010}.}
    \label{fig:wbexamples}
\end{figure}

For the example systems originally considered by \textcite{williams_nonnegative_2010} (their Figure 4, reproduced here in Figure~\ref{fig:wbexamples}) both PED approaches yield the same PID as $I_\text{ccs}$ \parencite{ince_measuring_2016} (Tables~\ref{tab:wbexamplesA},\ref{tab:wbexamplesB}).
Note that example A here is equivalent to the system \textsc{Subtle} in \textcite[][Figure 4]{griffith_intersection_2014}.
Only source redundancy is present in example A, with no mechanistic redundancy.

\begin{table}[htbp]
    \centering
    \begin{tabular}{| c | c | c | c | c | c |}
        \hline
        \textbf{Node} & $\bm{I_\partial} [ I_\text{min} ]$ & $\bm{I_\partial} [ I_\text{broja} ]$ & $\bm{I_\partial} [ I_\text{ccs} ]$ & $\bm{I_\partial} [ \textsc{ped} ]$ & $\bm{I_\partial} [ \text{\footnotesize mono-}\textsc{ped} ]$ \\
        \hline \hline
        $\{12\}$ &     $0.33$ & $0$ & $0.14$ & $0.14$ & $0.14$ \\ 
        $\{1\}$ &      $0.33$ & $0.67$ & $0.53$ & $0.53$ & $0.53$ \\
        $\{2\}$ &      $0.33$ & $0.67$ & $0.53$ & $0.53$ & $0.53$ \\ 
        $\{1\}\{2\}$ & $0.95$ & $0.25$ & $0.39$ & $0.39$ $(0.39,0)$ & $0.39$ $(0.39,0)$ \\ \hline
     \end{tabular}
     \caption{\emph{PIDs for system in Figure~\ref{fig:wbexamples}A.} (Figure 4A in \textcite{williams_nonnegative_2010}). Bracketed redundancy values decompose the redundant partial information into source and mechanistic redundancy respectively.}
  \label{tab:wbexamplesA}
\end{table}

\begin{table}[htbp]
    \centering
    \begin{tabular}{| c | c | c | c | c | c |}
        \hline
        \textbf{Node} & $\bm{I_\partial} [ I_\text{min} ]$ & $\bm{I_\partial} [ I_\text{broja} ]$ & $\bm{I_\partial} [ I_\text{ccs} ]$ & $\bm{I_\partial} [ \textsc{ped} ]$ & $\bm{I_\partial} [ \text{\footnotesize mono-}\textsc{ped} ]$ \\
        \hline \hline
        $\{12\}$ &     $0.5$ & $0$ & $0$ & $0$ & $0$ \\ 
        $\{1\}$ &      $0$ & $0.5$ & $0.5$ & $0.5$ & $0.5$ \\
        $\{2\}$ &      $0.5$ & $1$ & $1$ & $1$ & $1$ \\ 
        $\{1\}\{2\}$ & $0.5$ & $0$ & $0$ & $0$ $(0,0)$ & $0$ $(0,0)$ \\ \hline
     \end{tabular}
     \caption{\emph{PIDs for system in Figure~\ref{fig:wbexamples}B.} (Figure 4B in \textcite{williams_nonnegative_2010}). Bracketed redundancy values decompose the redundant partial information into source and mechanistic redundancy respectively.}
  \label{tab:wbexamplesB}
\end{table}

\subsection{Other examples}
\label{sec:otherex}
\textcite{griffith_quantifying_2014,griffith_intersection_2014} present a range of other interesting examples which by careful design exhibit different combinations of redundancy, synergy and unique information.

\begin{table}[htbp]
    \centering
    \begin{tabular}{| c  c  c || c |}
        \hline
        $x_1$ & $x_2$ & $s$ & $p(x_1, x_2, s)$ \\
        \hline \hline
        $0$ & $0$ & $0$ & $0.4$ \\
        $0$ & $1$ & $0$ & $0.1$ \\
        $1$ & $1$ & $1$ & $0.5$ \\ \hline
  \end{tabular}
  \caption{\emph{Definition of \textsc{ImperfectRdn}.}}
  \label{tab:uniquemisex}
\end{table}

\begin{table}[htbp]
    \centering
    \begin{tabular}{| c | c | c | c | c | c |}
        \hline
        \textbf{Node} & $\bm{I_\partial} [ I_\text{min} ]$ & $\bm{I_\partial} [ I_\text{broja} ]$ & $\bm{I_\partial} [ I_\text{ccs} ]$ & $\bm{I_\partial} [ \textsc{ped} ]$ & $\bm{I_\partial} [ \text{\footnotesize mono-}\textsc{ped} ]$ \\
        \hline \hline
        $\{12\}$ &     $0$ & $0$ & $0$ & $0.16$ & $0$ \\ 
        $\{1\}$ &      $0.39$ & $0.39$ & $0.23$ & $0.23$ & $0.23$ \\
        $\{2\}$ &      $0$ & $0$ & $0$ & $-0.16$ & $0$ \\ 
        $\{1\}\{2\}$ & $0.61$ & $0.61$ & $0.77$ & $0.77$ $(0.77,0)$ & $0.77$ $(0.77,0)$ \\ \hline
     \end{tabular}
     \caption{\emph{PIDs for \textsc{ImperfectRdn}}. \parencite[][Figure 3]{griffith_intersection_2014}. Bracketed redundancy values decompose the redundant partial information into source and mechanistic redundancy respectively.}
     \label{tab:rndunqxor}
\end{table}

We first consider \textsc{ImperfectRdn} \parencite[][Adapted from Fig. 3]{griffith_intersection_2014}.
The probability values for this system are shown in Table \ref{tab:uniquemisex} \parencite{ince_measuring_2016}.
This example is interesting because it is one for which $I_\text{ccs}$ reveals unique misinformation in the second predictor.
In the $I_\text{ccs}$ PID this was removed by thresholding negative values to $0$.
Here we see that the full PED approach produces a PID which is equal to the unthresholded $I_\text{ccs}$ decomposition, including the unique misinformation term, but that the monosemous PED approach which considers only unambiguous terms, has the same effect as thresholding.
It removes the counter-balanced effect of the term $H_\partial(\{12\}\{23\})$ which is both misinformation in the context of $I(X_2;X_3)$ and synergistic information in the context of $I(X_1,X_2;X_3)$.
Note that in this case, neither the $I_\text{ccs}$ or mono-\textsc{ped} approaches satisfy Eq.~\ref{eq:mi1pid}.

\begin{table}[htbp]
    \centering
    \begin{tabular}{| c | c | c | c | c | c |}
        \hline
        \textbf{Node} & $\bm{I_\partial} [ I_\text{min} ]$ & $\bm{I_\partial} [ I_\text{broja} ]$ & $\bm{I_\partial} [ I_\text{ccs} ]$ & $\bm{I_\partial} [ \textsc{ped} ]$ & $\bm{I_\partial} [ \text{\footnotesize mono-}\textsc{ped} ]$ \\
        \hline \hline
        $\{12\}$ &     $1$ & $1$ & $1$ & $2$ & $1$ \\ 
        $\{1\}$ &      $0$ & $0$ & $0$ & $-1$ & $0$ \\
        $\{2\}$ &      $0$ & $0$ & $0$ & $-1$ & $0$ \\ 
        $\{1\}\{2\}$ & $1$ & $1$ & $1$ & $2$ $(1,1)$ & $1$ $(1,0)$ \\ \hline
     \end{tabular}
     \caption{\emph{PIDs for \textsc{RdnXor}}. \parencite[][Figure 2]{griffith_intersection_2014}. Bracketed redundancy values decompose the redundant partial information into source and mechanistic redundancy respectively.}
     \label{tab:rndxor}
\end{table}

\textsc{RdnXor} consists of two two-bit (4 value) inputs $X_1$ and $X_2$ and a two-bit (4 value) output $S$.
The first component of $X_1$ and $X_2$ redundantly specifies the first component of $S$.
The second component of $S$ is the \textsc{xor} of the second components of $X_1$ and $X_2$. 
This system therefore contains $1$ bit of redundant information and $1$ bit of synergistic information; further every value $s \in S$ has both a redundant and synergistic contribution.
Table~\ref{tab:rndxor} shows the decompositions for this system. 
The monosemous PED approach agrees with the existing measures, while as for \textsc{xor} the full PED indicates $1$ bit of unique misinformation in each predictor, which comes from the $H_\partial(\{2\}\{12\})=H_\partial(\{1\}\{23\})=1$ terms. 
The monosemous PID correctly specifies the $1$ bit of redundancy as source redundancy.
The full PED also separates the $2$ bits of redundancy into $1$ bit of source (by construction of the system) and $1$ bit of mechanistic (arising from \textsc{xor}) redundancy.


\section{Discussion}

In this paper we start from a simple idea: to apply the framework of the Partial Information Decomposition \parencite{williams_nonnegative_2010} directly to multivariate entropy.
The concepts of synergy and redundancy can be applied to entropy in the same way they have been to mutual information.
Redundant entropy measures uncertainty that is shared or common between variables, while synergistic entropy measures extra uncertainty that arises when the variables are observed together, over and above that which is present in them individually.  
Similarly, the redundancy lattice of multivariate mutual information can be applied to entropy.
We define an entropy redundancy measure based on pointwise common surprisal which directly follows from a natural definition of shared entropy, satisfies the axioms required for validity of the redundancy lattice, and is closely related to the definition of mutual information.
Positive local information terms correspond to redundant entropy, negative local information terms to synergistic entropy. 
We show that this yields a consistent approach, which both addresses outstanding questions of how to meaningfully quantify dependence structure in multivariate systems \parencite{james_multivariate_2016} as well as suggesting new approaches to obtaining decompositions of multivariate mutual information. 

We suggest that the Partial Entropy Decomposition approach directly addresses the call made by \textcite{james_multivariate_2016} for new measures to meaningfully quantify the dependence structure of multivariate complex systems.
Here we have covered only the basic triadic and dyadic examples they present, showing that the PED can clearly reveal the different generative structures. 
Further applications of the PED directly to different model systems and real data is an interesting topic for future work. 
This problem has also been addressed by \textcite{rosas_understanding_2016}, where they employ an approach which symmetrises information redundancy by taking the minimum over all choices of the target variable. 

The PED, at least with the $H_\text{cs}$ measure, admits negative partial entropy terms.
Negative partial entropy values can occur in two different ways.
First, since $H_\text{cs}$ is not monotonic on the redundancy lattice \parencite{williams_nonnegative_2010}, this can result in negative partial entropy values (e.g. \textsc{and}, Sec.~\ref{sec:and}).
Second, even in cases where the redundant entropy values are monotonic on the lattice, the particular statistical structure of a system can also induce negative partial entropy values (e.g. \textsc{xor}, Sec.~\ref{sec:xor}; \textcite{rauh_reconsidering_2014-1}).
We agree with \textcite{james_multivariate_2016} that while ``negative \dots~atoms can \emph{subjectively} be seen as flaw'' the ability to meaningfully distinguish the generative structure of systems as we can with the PED is more important than the aesthetic benefits of a non-negative decomposition.
We suggest these negative terms are actually an important part of the structure of the multivariate system.
For example, we have shown that negative pairwise entropy redundancy between two variables interpreted as predictors in the PID framework, quantifies mechanistic information redundancy in the system.
Similarly, the 3-way entropy redundancy between pairwise synergies, when negative, is likely to indicate the presence of mechanistic pairwise redundancy.
Whether other terms can take negative values, and whether there are similar interpretations of such in terms of mechanistic redundancy is an open question.

The relationships between mutual information values and the partial entropy decomposition presented here are valid for any redundancy measures.
It is therefore possible that alternative measures to obtain partial entropy values \parencite{bertschinger_quantifying_2014}, or alternative perspectives on the redundancy lattice structure \parencite{chicharro_synergy_2017}, or modifications to the definition of $H_\text{cs}$ perhaps involving the use of different distributions for pairwise terms (e.g. $P_\text{ind}$, \textcite{ince_measuring_2016}) might yield a non-negative entropy decomposition.
However, it is unclear then whether such methods would be able to quantify mechanistic redundancy. 
Further, the close relationship between $H_\text{cs}$ and the definition of mutual information is appealing.

In this manuscript we focus on the implications of the PED perspective for obtaining decompositions of mutual information.
Mutual information is based on entropy, and the concepts of synergy and redundancy apply equally to entropy. 
Therefore individual partial information terms in any valid PID must be quantifying combinations of redundant and synergistic entropy.
Hence, we suggest that every consistent PID should admit a compatible entropy decomposition. 

Considering redundant and synergistic entropy directly reveals the insight that local misinformation \parencite{wibral_bits_2014}, which has been considered difficult to interpret \parencite{williams_nonnegative_2010}\footnote{``it is entirely unclear what it means for one variable to provide `negative information' about another''} is synergistic entropy.
This also supports the definition of $H_\text{cs}$ for which this equivalence is true locally.
The relationships between mutual information and partial entropy terms are fixed by the structure of the lattices, independent of the redundancy measure used (Sec.~\ref{sec:mifromped}). 
Therefore, considering Eq.~\ref{eq:mi_ped} it is hard to imagine a different entropy redundancy function that could yield a non-negative decomposition but that fails to match the pointwise definition of mutual information. 

Equation \ref{eq:mi_ped} also provides the insight that mutual information between two variables includes both redundant and synergistic effects, and this realisation has important consequences related to the definitions of the PID, for example, the identity axiom (Section \ref{sec:identity}). 
Similarly, it suggests that other axioms relating partial information terms to mutual information, such as the  bivariate monotonity PI axiom \parencite{bertschinger_quantifying_2014} should be reconsidered\footnote{Rather than $I_\partial(3;\{1\}\{2\}) \leq I(X_1; X_2)$ we suggest the redundant partial information should include only the redundant entropy effects, not the synergistic entropy (misinformation), resulting in bivariate monotonicity being defined as $I_\partial(3;\{1\}\{2\}) \leq H_\partial^{12}(\{1\}\{2\})$.}. 
It also gives further support to our claim that any partial information decomposition must admit negative terms \parencite{ince_measuring_2016}, corresponding, for example, to unique misinformation which is synergistic entropy that arises only when that predictor and the target are considered together (e.g. \textsc{ImperfectRdn}, Sec.~\ref{sec:otherex}) or mechanistic redundancy in a system when the target variable is switched (e.g. \textsc{and}, Sec~\ref{sec:and}).
While mutual information over all is guaranteed to be non-negative, there is no such guarantee for the unique contributions from individual variables or combinations of variables, which do not have a corresponding classical Shannon quantity.

By considering the relationship between the PED and multivariate mutual information we derive two alternative information decompositions. 
An important insight that arises from this process is the fact that, in entropy terms, $I(X_1;S)$ is not a proper subset of $I(X_1,X_2;S)$; the former includes partial entropy terms that do not appear in the latter.
This fact, which has not been recognised before, seems to be a key factor explaining why obtaining a consistent PID has been challenging.

The first PED based information decomposition satisfies all the standard properties of the PID, including Eqs.~\ref{eq:mi12pid},\ref{eq:mi1pid}.
However, this provides quite a different perspective to existing approaches in several examples (e.g. \textsc{xor}, Sec.~\ref{sec:xor}).
These differences arise from inherently ambiguous entropy terms, that can correspond for example both to unique misinformation in the context of a univariate predictor mutual information as well as synergistic information in the context of the full bivariate predictor mutual information. 
These are the terms which appear in the univariate mutual information, but not in the joint mutual information, and are only accessible from the entropy perspective; they can not be separated when considering mutual information values directly. 
However, since these ambiguous terms do not actually occur directly in the bivariate predictor mutual information this suggests a second information decomposition we term the monosemous decomposition. 
While this no longer satisfies Eq.~\ref{eq:mi1pid}, it is nevertheless a valid decomposition of the joint mutual information into components which are unambiguously redundant, unique or synergistic between the predictor variables.
In fact, it provides a genuinely new perspective with strikingly different results to existing methods in simple example systems. 
We suggest this alternative approach may be particularly useful for gaining insight into the functional structure of a system, since the decomposition for simple binary operations seems to nicely reflect their generative mechanism (Sec.~\ref{sec:binsummary}). 

By considering a measure of dependence based only on redundant entropy (Sec.~\ref{sec:puremi}), we illustrated directly how the difficulties with the PID come from the inclusion of synergistic effects in mutual information.
In the context of this ``pure mutual information'' measure, it is trivial to obtain a decomposition which satisfies all the relationships of the PID.
In all the examples considered here, this approach is numerically equivalent to the monosemous PED.
This serves as an exercise to demonstrate that although the mono-PED does not satisfy Eq.~\ref{eq:mi1pid}, there is a perspective (that of decomposing only shared entropy), in which a similar approach can satisfy such properties.
Pure mutual information also highlights the origin of the many failures of intuition regarding axioms and properties of the PID.
Most of these apply as previously proposed to pure mutual information, but are incorrect as defined for mutual information because of the hitherto under appreciated synergistic entropy component.

The PED approach also allows separate quantification of mechanistic vs source redundancy \parencite{harder_bivariate_2013} for both PED based decompositions, even when both types of redundancy occur together in the same system (e.g. \textsc{and}, Sec.~\ref{sec:and}).
Despite the importance of this distinction to separate properties of functional mechanisms from dependencies between input variables, to our knowledge this is the first proposal for quantifying these differences in practise. 

Both the PID from PED approaches are derived from a single entropy lattice, so they are consistent with respect to a choice of target, in a way that existing PID measures are not.
Neither are non-negative for the reasons discussed above. 
However, it may be possible to obtain some theoretical bounds, and this is an interesting area for future work.
For example, given that mutual information is non-negative, in the two variable redundancy lattice redundant entropy must be greater than synergistic entropy\footnote{$H_\partial(\{1\}\{2\}) \geq  H_\partial(\{12\})$, with equality only if $H_\partial(\{1\}\{2\})=H_\partial(\{12\})=0$.}.
Similar multi-term constraints can be obtained from the PED expressions for the individual and joint mutual informations. 

We note that $H_\text{cs}$ based entropy decompositions are mostly consistent with the $I_\text{ccs}$ based PID \parencite{ince_measuring_2016}.
Where there are differences, these are induced from the use of $P_\text{ind}$ in $I_\text{ccs}$ which induces the asymmetry in different PIDs of the same multivariate system.

It will be important to extend the PED approach to larger numbers of variables, for example a four variable PED yielding three variable PIDs. 
The relationship between the lattices and hence the PED expressions for different mutual information terms should be straightforward to formalise and we expect the approach here to apply directly. 
One question concerns the use of the marginally constrained maximum entropy distribution for four source redundant entropy terms. 
Should pairwise constraints continue to be used --- this is closer to the arguments for the PID that redundancy should depend only on pairwise target-predictor marginals and no higher order structure \parencite{bertschinger_quantifying_2014} --- or should third order marginal constraints be used.
Full development and testing on higher dimensional example systems \parencite{griffith_quantifying_2014,griffith_intersection_2014,ince_measuring_2016} is an important area for future work. 

The definition of $H_\text{cs}$ is easily applicable to continuous systems \parencite{ince_measuring_2016}.
Future work will implement the entropy decomposition for Gaussian systems, which offer a number of practical advantages \parencite{barrett_exploration_2015,ince_statistical_2016-1}. 

In conclusion, we suggest that the problem of quantifying the structure of multivariate dependence is aided by a simple perspective shift from considering mutual information to focus instead on entropy.
Entropy is the foundational, first order quantity of information theory, and so it seems a more natural place to start for concepts which have proved difficult to formalise, such as redundancy and synergy. 
The elegant framework of the Partial Information Decomposition can be naturally and directly applied to multivariate entropy.
We believe this perspective shift helps to clarify some of the difficulties around information theoretic quantification of multivariate systems, in terms of both entropy \parencite{james_multivariate_2016} and mutual information \parencite{williams_nonnegative_2010}.
We hope the resulting entropy and information decompositions will help to clarify theoretical aspects as well as providing useful tools for practical data analysis in the study of complex systems and experimental science. 

\section*{Acknowledgements}

I thank Bill Phillips, Daniel Chicharro, Johannes Rauh, Ryan James and Philippe Schyns for useful discussions, and Ryan James for producing the excellent \texttt{dit} package.

\printbibliography

\end{document}